\documentclass[twocolumn,amssymb, nobibnotes, aps, prb]{revtex4-2}
\usepackage[bookmarks=false]{hyperref}
\usepackage[utf8]{inputenc}
\usepackage[english]{babel}
\usepackage{amssymb}
\usepackage{amsmath}
\usepackage{graphicx}
\usepackage{color}
\usepackage{physics}
\usepackage{float}
\renewcommand{\vec}[1]{\boldsymbol{#1}}
\graphicspath{{Figs/}}
\newcommand{\be}{\begin{equation}}
\newcommand{\ee}{\end{equation}}
\newcommand{\bea}{\begin{eqnarray}}
\newcommand{\eea}{\end{eqnarray}}

\def\nn{\nonumber}
\def\lb{\label}
\def\pref{\eqref}

\def\g{\gamma}

\def\tt{\hat \tau}

\begin{document}

\title{
Band structure and insulating states driven by the Coulomb interaction in twisted bilayer graphene
}

\author{Tommaso Cea$^1$}
\author{Francisco Guinea$^{1,2}$}
\affiliation{$^1$Imdea Nanoscience, Faraday 9, 28015 Madrid, Spain}
\affiliation{$^2$ Donostia International Physics Center, Paseo Manuel de Lardizábal 4, 20018 San Sebastián, Spain}

\date{\today}

\begin{abstract}
We analyze the phase diagram of twisted graphene bilayers near a magic angle. We consider the effect of the long range Coulomb interaction, treated within the self consistent Hartree-Fock approximation, and we study arbitrary band fillings. We find a rich phase diagram, with different broken symmetry phases, although tehy do not show necessarily a gap at the Fermi energy. There are non trivial effects of the electrostatic potential on the shape and the gaps of the bands in the broken symmetry phases. The results suggest that the non superconducting broken symmetry phases observed experimentally are induced by the long range exchange interaction.
\end{abstract}

\maketitle
Twisted bilayer graphene (TBG) near the "magic angles"\cite{Bistritzer_pnas11} shows a rich phase diagram\cite{cao_nat18,cao_fatemi_nat18,Yankowitz_science19,lu_nat19} with a variety of insulating and superconducting phases. While the existence of superconductivity is well established, the number, and nature of the insulating phases is still only partially understood.

Simple order of magnitude arguments show that the leading electron-electron interaction in TBG near the magic angles is the long range Coulomb interaction. The strength of this interaction can be estimated to be: $E_C \sim e^2 / ( \epsilon L ) \approx 10 - 15$ meV, where $\epsilon \approx 8 - 10$ is the screening from the (mostly hBN) environment, and $L \sim 12 - 15$ nm is the length of the moiré lattice unit. This energy scale is larger than the bandwidth at the magic angles, $W \lesssim 5$ meV. 

The standard way to treat long range electrostatic interactions in condensed matter physics is by using the self consistent Hartree-Fock approximation. This approach takes into account the leading effect of the screened electrostatic potential. The spin and valley degeneracy of the non interacting system implies that the Hartree term, which includes interactions of each electron species with all the others, is dominant. The Fock term, which is spin and valley dependent, allows for a variety of broken symmetry phases, although it cannot describe superconductivity. Note, however, that the analysis of superconducting phases requires a knowledge of the electronic structure and of the shape of the Fermi surface. Similarly, the study of possible fractional Chern insulator phases\cite{RS19,LTKV19,ALB20} is outside the scope of this work, although the understanding of these phases needs as an input the electronic properties reported here. A number of theoretical works have analyzed broken symmetry phases for specific fillings\cite{GLGS17,Letal18,SB18,KLK18,po_prx18,zou_prb18,bultinck_cm19,liu_cm19,GS19,LQYZLH19,cea_prb19,NIL19,IF19,FV19,BJ19,CCC19,SRRN19,dai_cm19,zhang_cm20,gonzalez_cm20,KV19,SKU19,XM20,gonzalez_cm20,kang_cm20,HWS20,KKH20}. We will compare our findings to these analyses in the following.

We present results for the electronic structure and the stability of a number of broken symmetry phases for arbitrary fillings of TBG near the first magic angle, $\theta\simeq 1.05^\circ$. Our results suggest that the insulating phases found experimentally can be understood within the Hartree-Fock approximation. The interplay between purely electrostatic (Hartree) and exchange (Fock) effects makes these phases quite unlike insulating polarized phases discussed in other contexts in condensed matter physics.

We study the non interacting electronic structure of TBG using the continuum model\cite{santos_prl07} obtained by combining the Dirac equations from the two layers (see\cite{si}).
The self consistent Hartree-Fock approximation can be seen as a variational approach where an effective potential is defined such that the ground state (GS) wavefunction that it leads to is the best Slater determinant wavefunction for the original Hamiltonian. The effective potential includes a purely electrostatic (Hartree) term, which gives the average effect of each electron on all the others, and an exchange term, which only couples electrons with the same internal quantum numbers, spin and valley. The Hartree term is described by a potential, $V_H ( \vec{r} ) = \sum V_H\left({\vec{G}_i}\right) e^{i \vec{G}_i\cdot \vec{r}}$, which is local in real space. It has the symmetry of the moiré lattice, and it can be expanded in a Fourier series determined by reciprocal lattice vectors, $\vec{G}_i$. Previous calculations\cite{cea_prb19,Guinea_pnas18} show that this expansion converges rapidly, and only the six leading reciprocal vectors, $| \vec{G}_i | = ( 4 \pi ) / ( \sqrt{3} L ) , i = 1 , \cdots , 6$ are needed. We assume that the Hartree potential is such that it is exactly cancelled by the other charges in the system at the charge neutrality (CN) point of the non interacting system, which is consistent with tight binding calculations\cite{rademaker_prb19}. 

The Fock potential is non local in real space. It leads to a self energy, $\Sigma_F ( \vec{k} + \vec{G} , \vec{k} + \vec{G}' )$, which involves, in turn, a summation over occupied states, and additional momentum transfers\cite{si}:
\bea
   \Sigma_F^\mu ( \vec{k} + \vec{G} , i ; \vec{k} + \vec{G}' , j )& =&-\Omega^{-1} \sum_{\vec{k}' , \vec{G}'' , \alpha} v_C ( | \vec{k} - \vec{k}' - \vec{G}'' | ) \times \nn \\
   &\times& \phi_{\vec{k}' + \vec{G} + \vec{G}'' , \alpha , \mu , i} \phi^*_{\vec{k}' + \vec{G}' + \vec{G}'' , \alpha , \mu , j} 
   \label{fock}
\eea
where $\Omega$ is the volume, $\phi_{\vec{k} + \vec{G} , \alpha , i}$ is the amplitude of the state in the band $\alpha$ on a wavefunction with momentum $\vec{k} + \vec{G}$, $\vec{k}$ belongs to the reduced Brillouin zone (BZ), and sublattice/layer indices $i , j = 1 , \cdots , 4$. The label $\mu = 1, \cdots , 4$ stands for valley and spin. Finally, the Fourier transform of the Coulomb potential is
\begin{align}
    v_C \left(  \vec{q}\right  ) &= \frac{2 \pi e^2}{ \epsilon | \vec{q} |} \tanh ( | \vec{q} | d )
    \label{coulomb}
\end{align}
where $e$ is the electron charge, $\epsilon = 10$ is the dielectric constant of the environment, and $d = 40$ nm is the distance between to the metallic gates placed at the same distance above and below the TBG.

The sum in the Eq. \pref{fock} runs over occupied bands $\alpha$ and momenta $\vec{k}$. We assume that the contribution to the Fock potential from bands other than the two ones closest to CN do not change as the occupancy of these narrow bands is modified. Hence, we assume that the exchange potential arising from these bands only contributes to the value of the Fermi velocity in the non interacting Hamiltonian. We have checked that the summation over the reciprocal lattice vectors $\vec{G}$ in the Eq. \pref{fock} converges rapidly after the first set of six vectors. It's worth notice that the exchange self energy is diagonal in valley and spin. 

Broken symmetry phases in the Hartree-Fock approximation are described by exchange self energies which either break spatial symmetries present in the non interacting Hamiltonian, or break the equivalence between the electron flavors, valley and spin. The last case implies the absence of symmetry under time reversal. We consider i) phases which break the equivalence between the two sublattices within each layer, what makes possible a gap at the Dirac points of the non interacting Hamiltonian\cite{c2}, and ii) phases where the exchange potential depends on the spin or valley of the electron, but which conserve the spatial symmetries of the non interacting Hamiltonian.

In the following we adopt the parametrization of the twisted bilayer graphene
given in the Ref.\cite{koshino_prx18}: $\hbar v_F/a=2.1354$eV,
$a=2.46${\AA},
$g_1=0.0797$eV and $g_2=0.0975$eV.
The difference between $g_1$ and $g_2$, as described in the Ref.\cite{koshino_prx18},
accounts for the corrugation effects where the interlayer distance is minimum at the $AB/BA$ spots and maximum at $AA$ ones, or it can be seen as the outcome of a more complete treatment of the lattice relaxation\cite{guinea_prb19}.

\begin{figure}
\centering
\includegraphics[width=3.in]{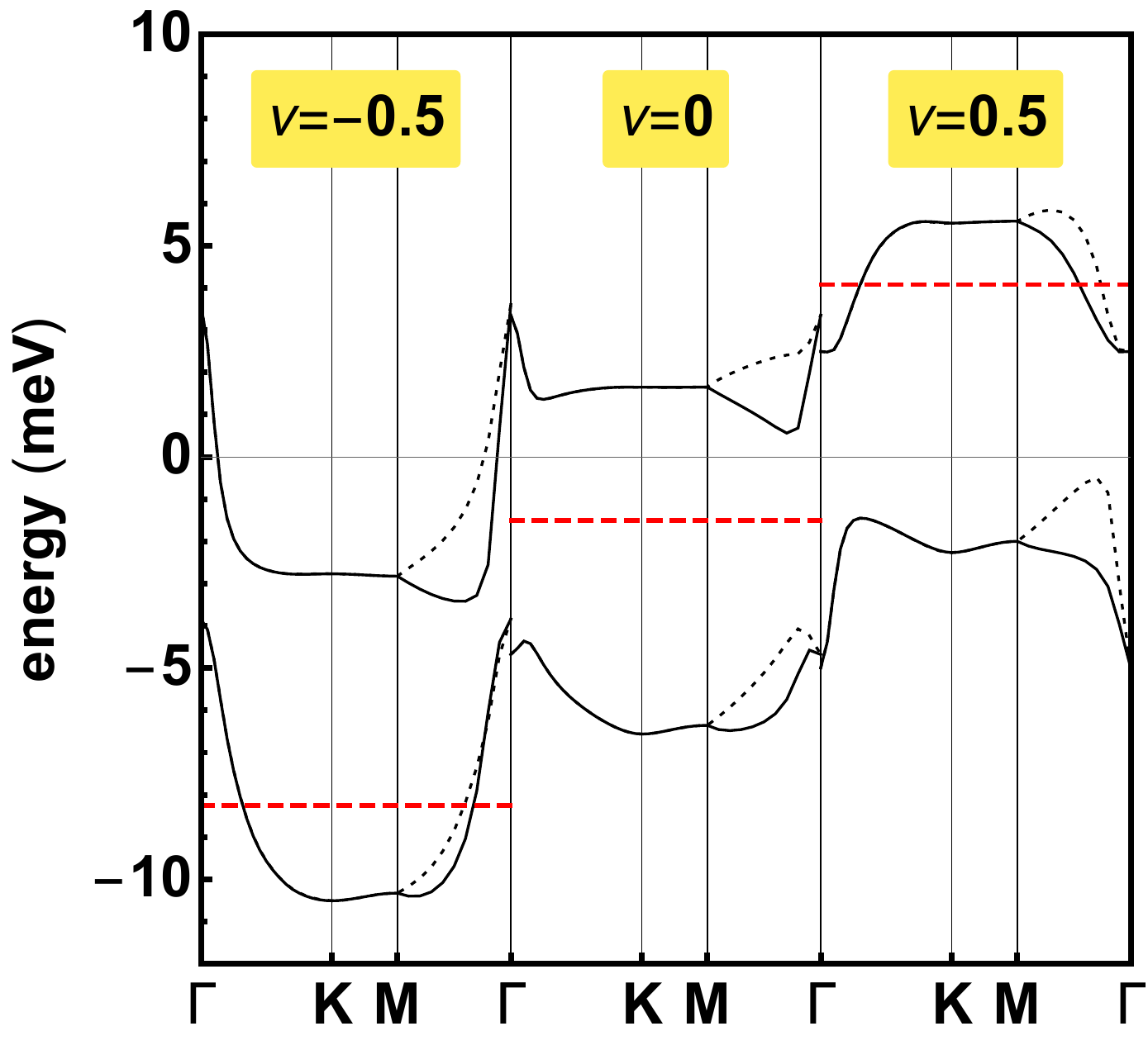}
\caption{Bands for a phase which breaks $\mathcal{C}_2 \mathcal{T}$ symmetry and opens a gap between the conduction and valence bands. Screening of the Coulomb potential is described by a dielectric constant $\epsilon = 10$. This self consistent solution exists near half filling.}
\label{np_bands_evolution}
\end{figure}

\begin{figure*}
\centering
\includegraphics[width=3.5in]{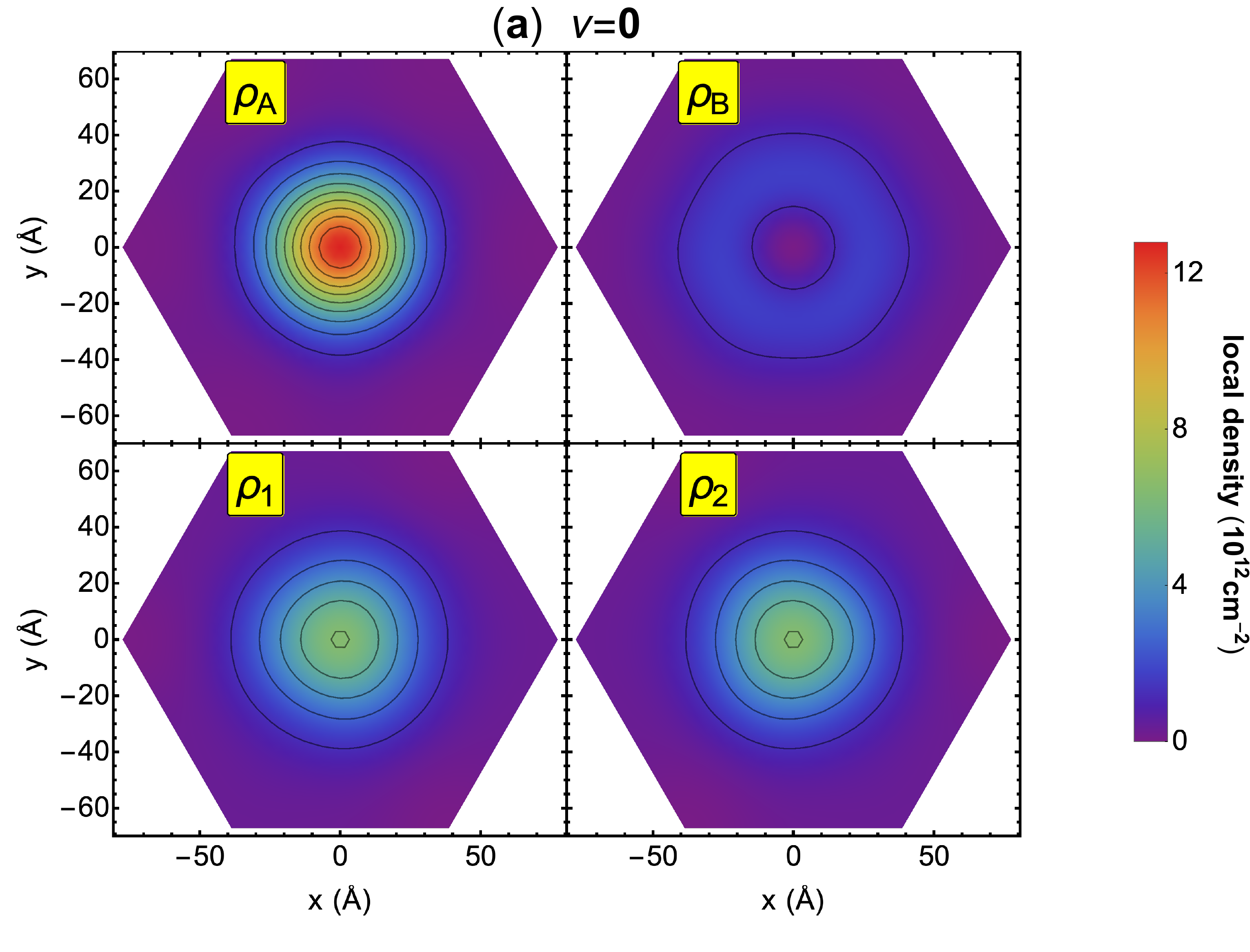}
\includegraphics[width=3.5in]{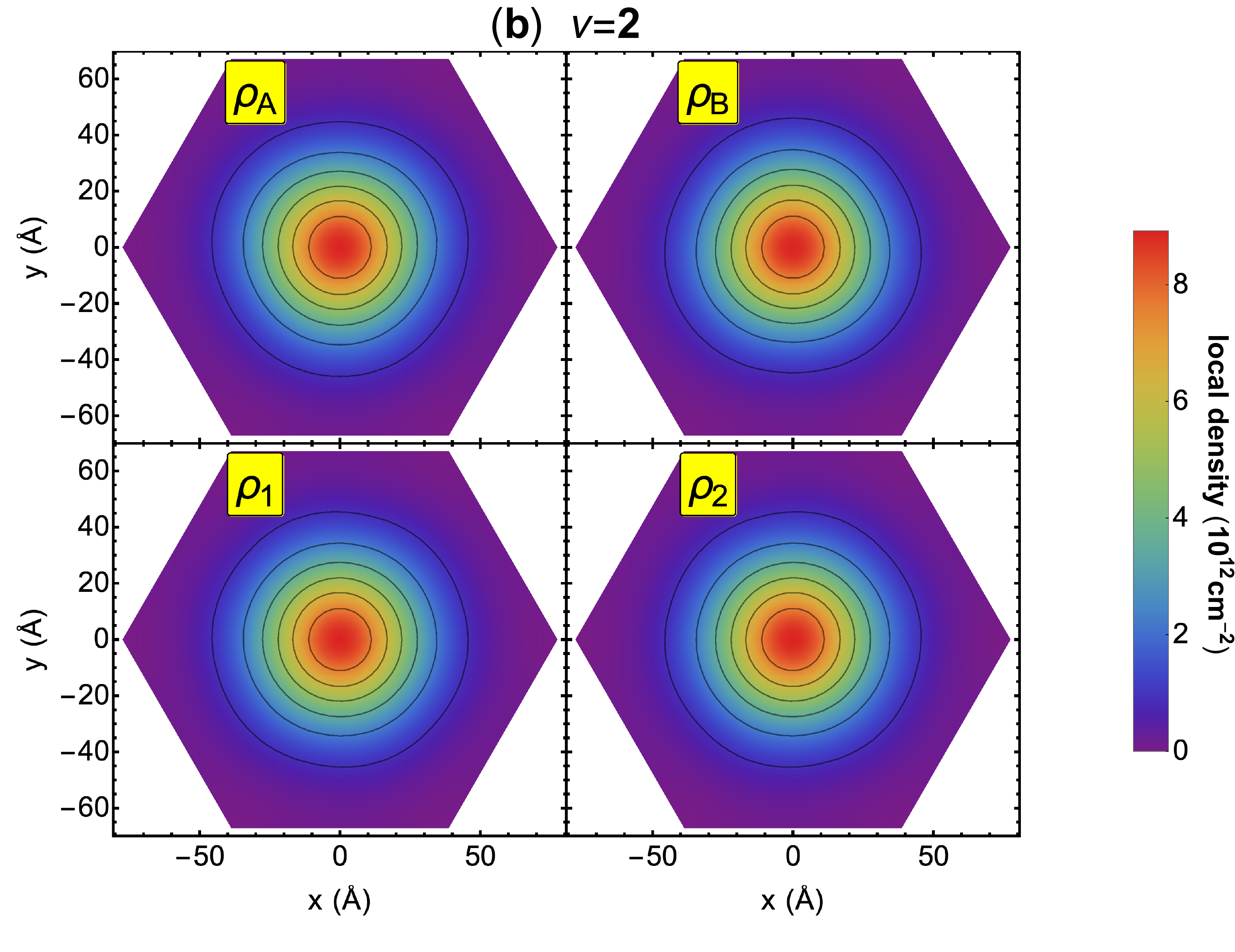}
\caption{
Charge density distribution obtained for $\nu=0$ (a) and $\nu=2$ (b).
$\rho_{A,B}$ is the local charge computed in the sub-lattice $A,B$, respectively, while $\rho_{1,2}$ is that corresponding to the layer $1,2$, respectively.
}
\label{charge}
\end{figure*}

We consider first solutions where the spatial symmetry of the non interacting Hamiltonian is broken, but where there is no spin or valley polarization. In order to achieve these solutions, a small symmetry breaking term is introduced at the beginning of the interactions towards self consistency. The bands obtained in this way are shown in the Fig.[\ref{np_bands_evolution}]. We obtain self consistent solutions of this type at CN, and for filling factors $-1 \lesssim \nu \lesssim 1$. Solutions with broken $\mathcal{C}_2$ symmetry cease to be stable for fillings $\nu = \pm 1.5$ (see\cite{si}). Outside this density range the Hartree term dominates, and increases the bandwidth\cite{Guinea_pnas18,rademaker_prb19,cea_prb19}. The self consistent broken symmetry solution combines four equivalent wavefunctions, one for each set of valley and spin indices. The order parameter is the difference between the charge density at the $A$ and $B$ sublattices in both layers, as shown in the Fig.[\ref{charge}]. Note that the two layers show very similar charge densities. The relative sign of the order parameter in each of the four spin and valley sectors cannot be estimated from the effect of long range interactions only\cite{si}. It is likely that short range interactions will favor a phase where the two sublattices are equally occupied, due to order parameters with different signs for different flavors, as in the broken symmetry phases in graphene where the $n=0$ Landau level is partially occupied\cite{AF06}. A gapped phase, with broken $\mathcal{C}_2 \mathcal{T}$ symmetry, has been discussed in the Refs.\cite{bultinck_cm19,liu_cm19,XM20,gonzalez_cm20}. The gaps reported in\cite{bultinck_cm19,liu_cm19} are larger than those shown in the Fig.[\ref{np_bands_evolution}], most likely due to the use of a lower dielectric constant.  The energy of this phase will be later compared to the energies of other broken symmetry phases.

\begin{table}[h!]
\begin{center}
\begin{tabular}{|l|c|c|r|}
\hline
& $2+2$ & $1+3$ & $3+1$ \\
\hline
$n_-$ & $4+\frac{\nu-|\nu|}{2}$&$3+\frac{\nu-|\nu+2|}{2}$ & $5+\frac{\nu-|\nu-2|}{2}$\\
\hline
$n_+$ & $\frac{\nu+|\nu|}{2}$&1+$\frac{\nu+|\nu+2|}{2}$ & $-1+\frac{\nu+|\nu-2|}{2}$\\
\hline
\end{tabular}
\end{center}
\caption{Occupation number of the low (high) occupancy bands, $n_-$($n_+$), corresponding to the configurations of the GS specified in the upper panels.}\label{polarization_filling_table}
\end{table}

\begin{figure*}
\centering
\includegraphics[width=2.5in]{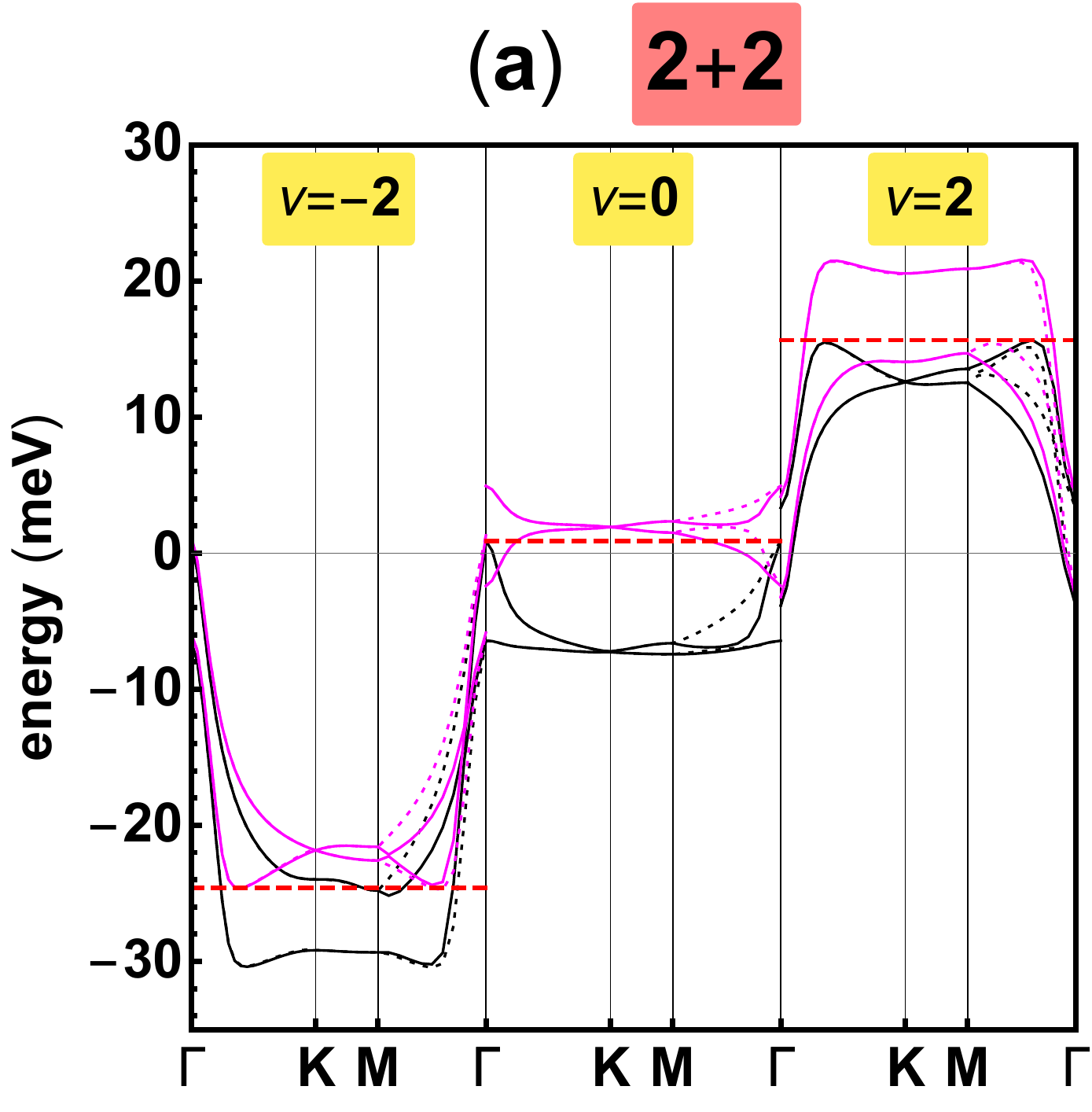}
\includegraphics[width=2.5in]{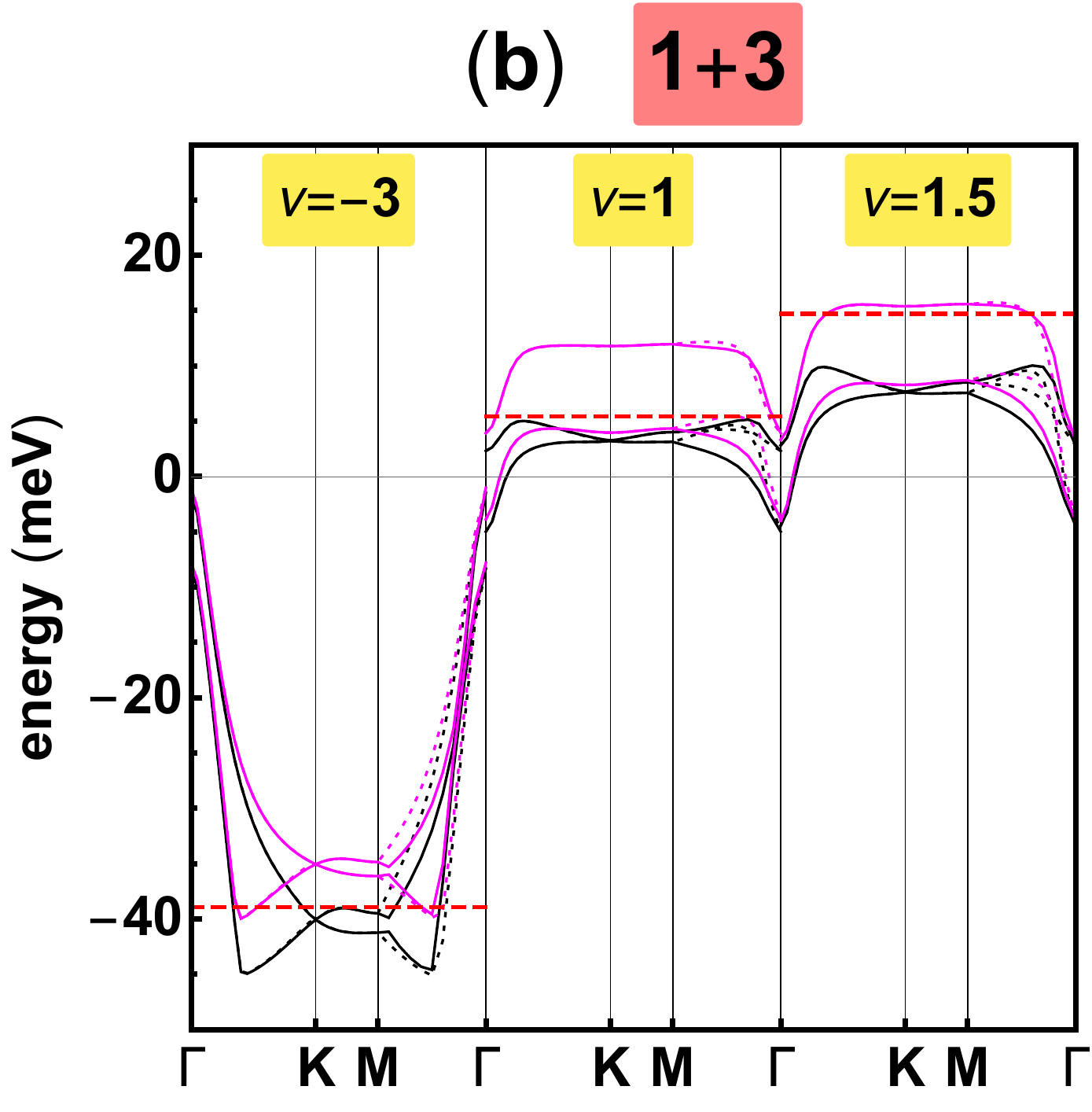}
\includegraphics[width=2.5in]{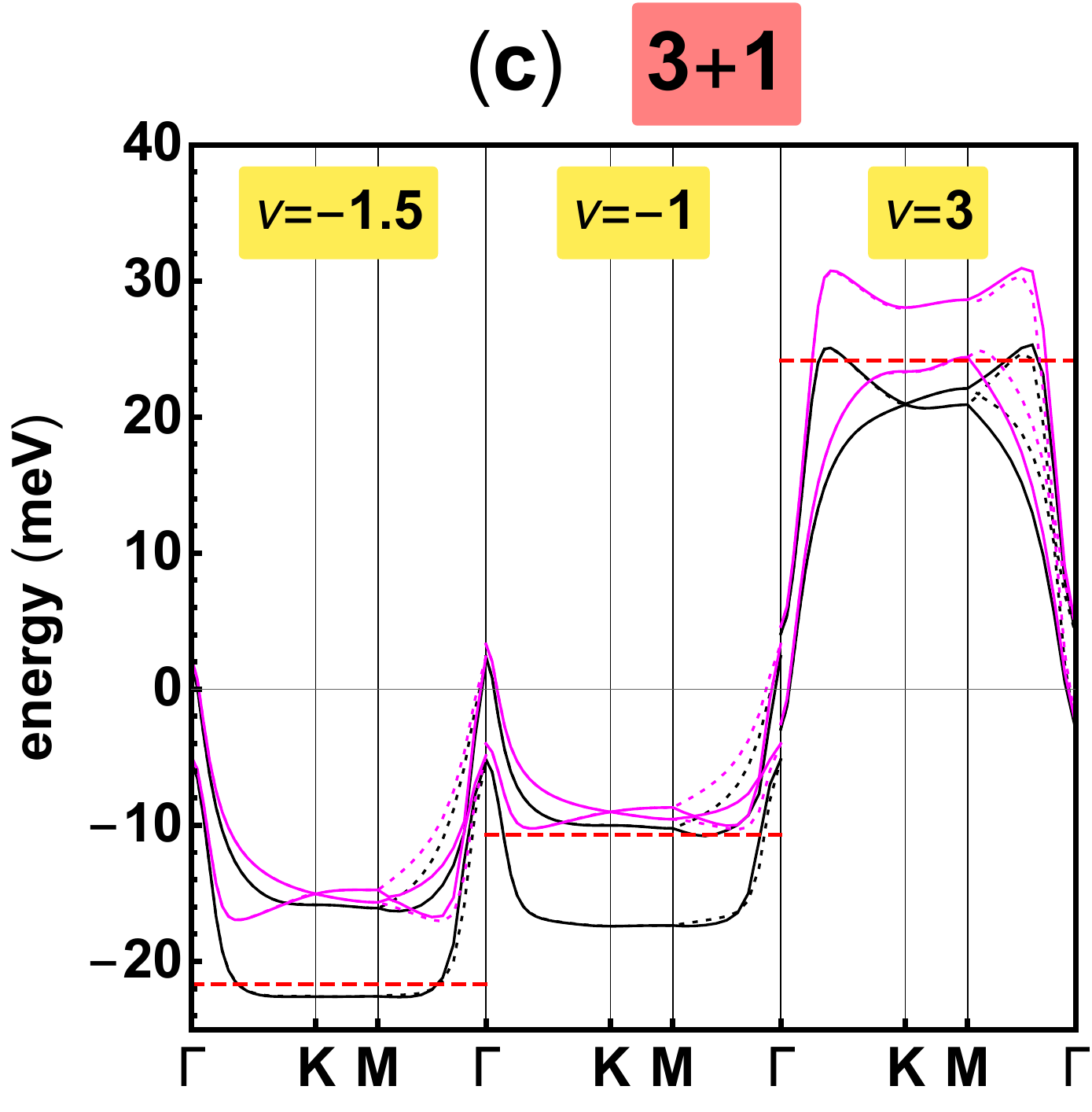}
\caption{
Band structure obtained at integer fillings, $\nu\in[-3,3]$,
for the polarized configurations: $2+2$ (a), $1+3$ (b) and $3+1$ (c).
}
\label{polarized_bands_evolution}
\end{figure*}

\begin{figure*}
\begin{tabular}{c} 
\includegraphics[width=3in]{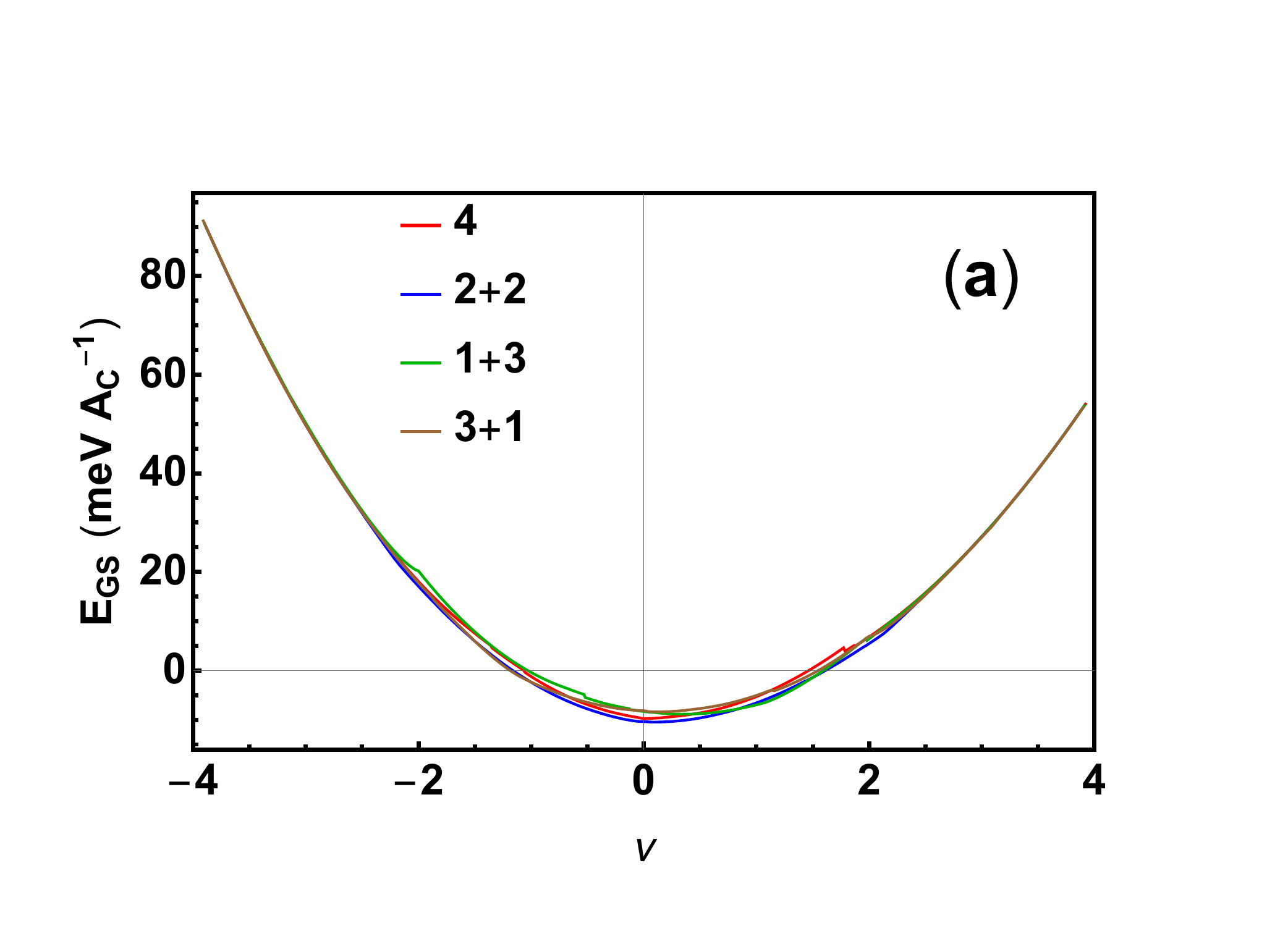} \\
\includegraphics[width=3in]{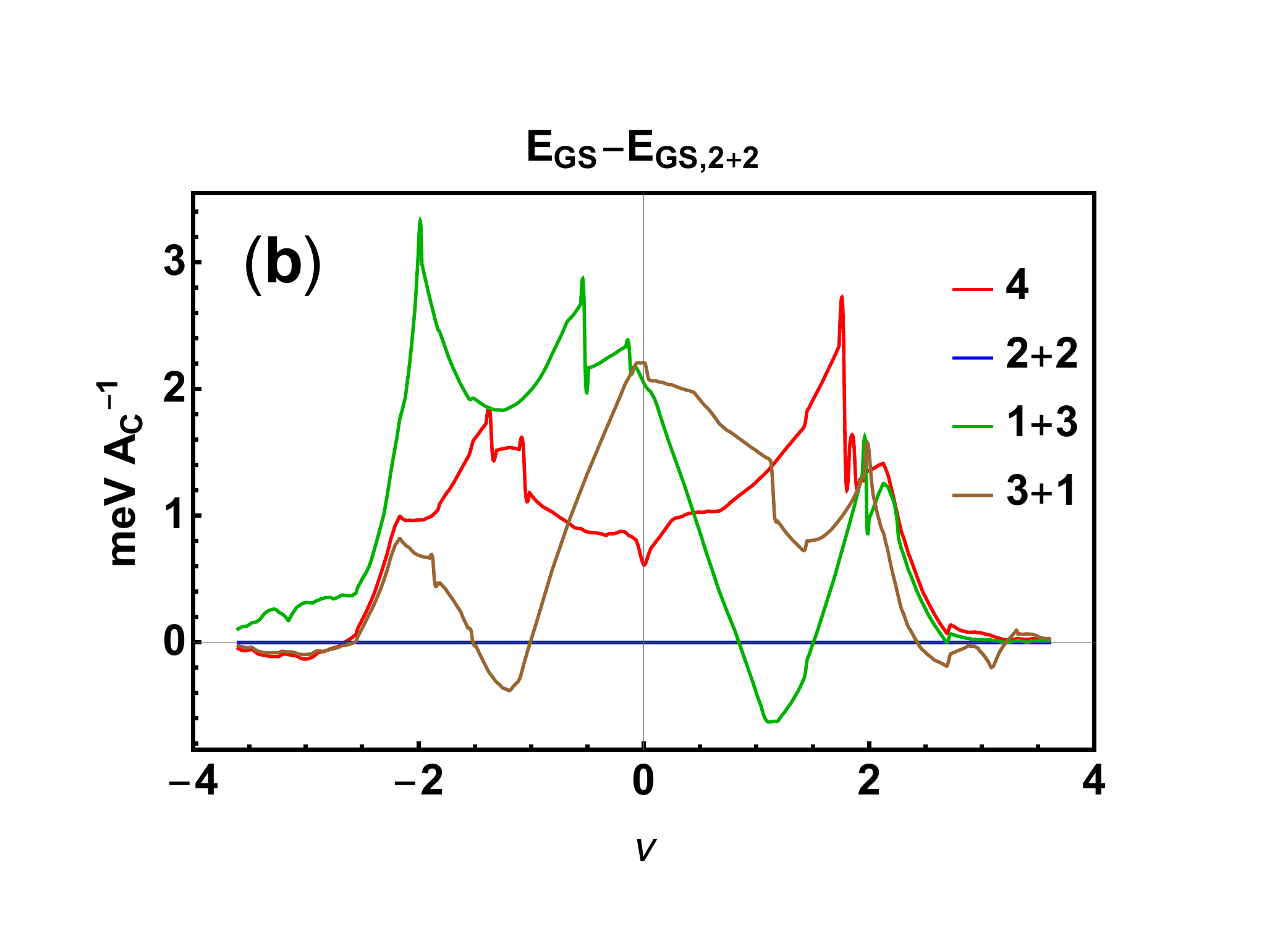}
\end{tabular}
\caption{(a): energy of the GS per moir\'e unit cell, computed for the four possible configurations. The label $4$ refers to the non-polarized GS. $A_C=\sqrt{3}L^2/2$ is the area of the unit cell.
(b): difference between the energy of the GS in each configuration and that corresponding to the configuration $2+2$.}
\label{EGS}
\end{figure*}

\begin{figure*}
\centering
\includegraphics[width=3in]{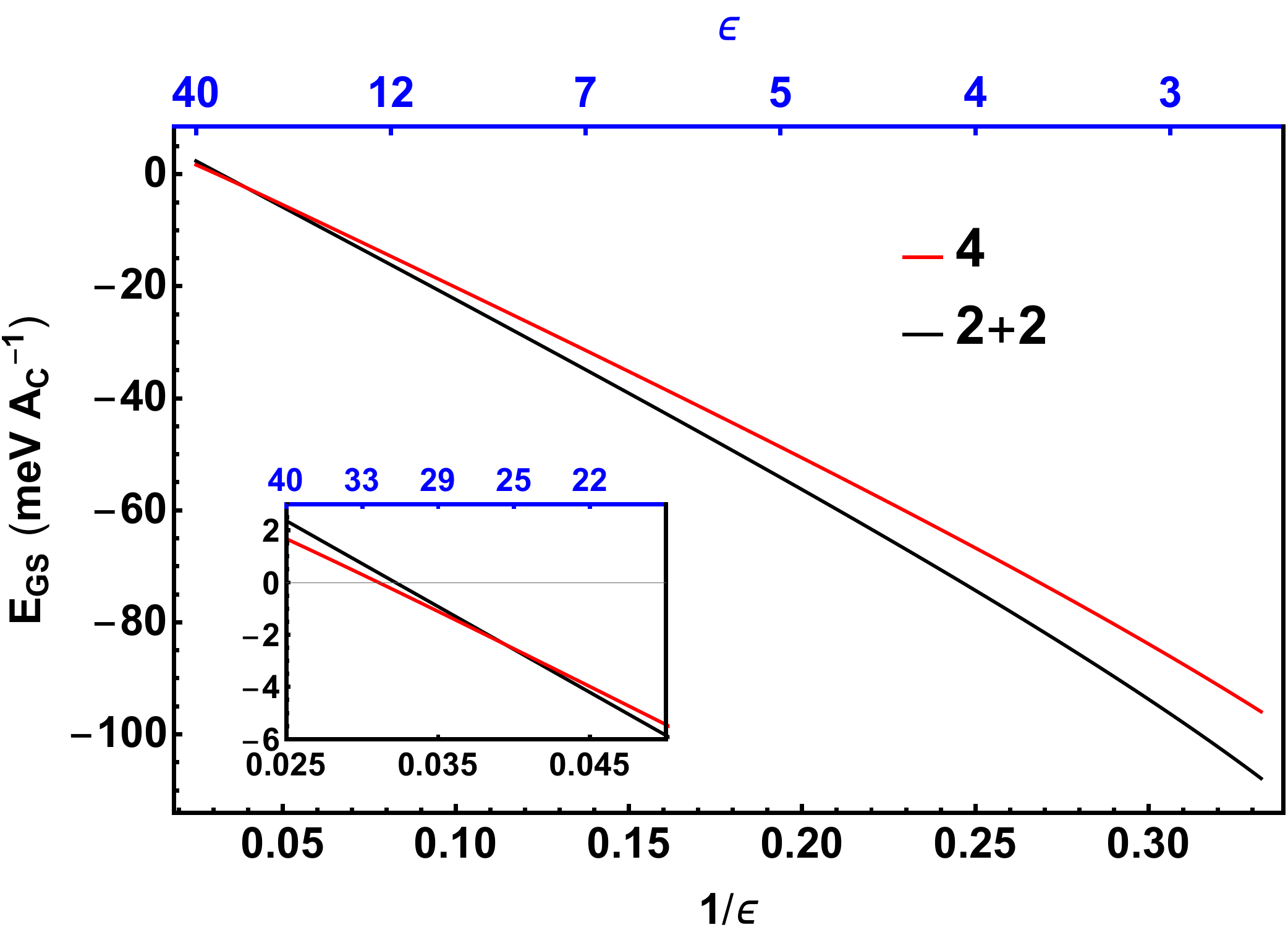}
\caption{Comparison of he GS energies per moir\'e unit cell for the phase with broken $\mathcal{C}_2 \mathcal{T}$ symmetry and the polarized 2+2 phase at half filling, as function of the dielectric constant, $\epsilon$.}
\label{EGS_epsilon}
\end{figure*}

\begin{figure*}
\centering
\includegraphics[width=5in]{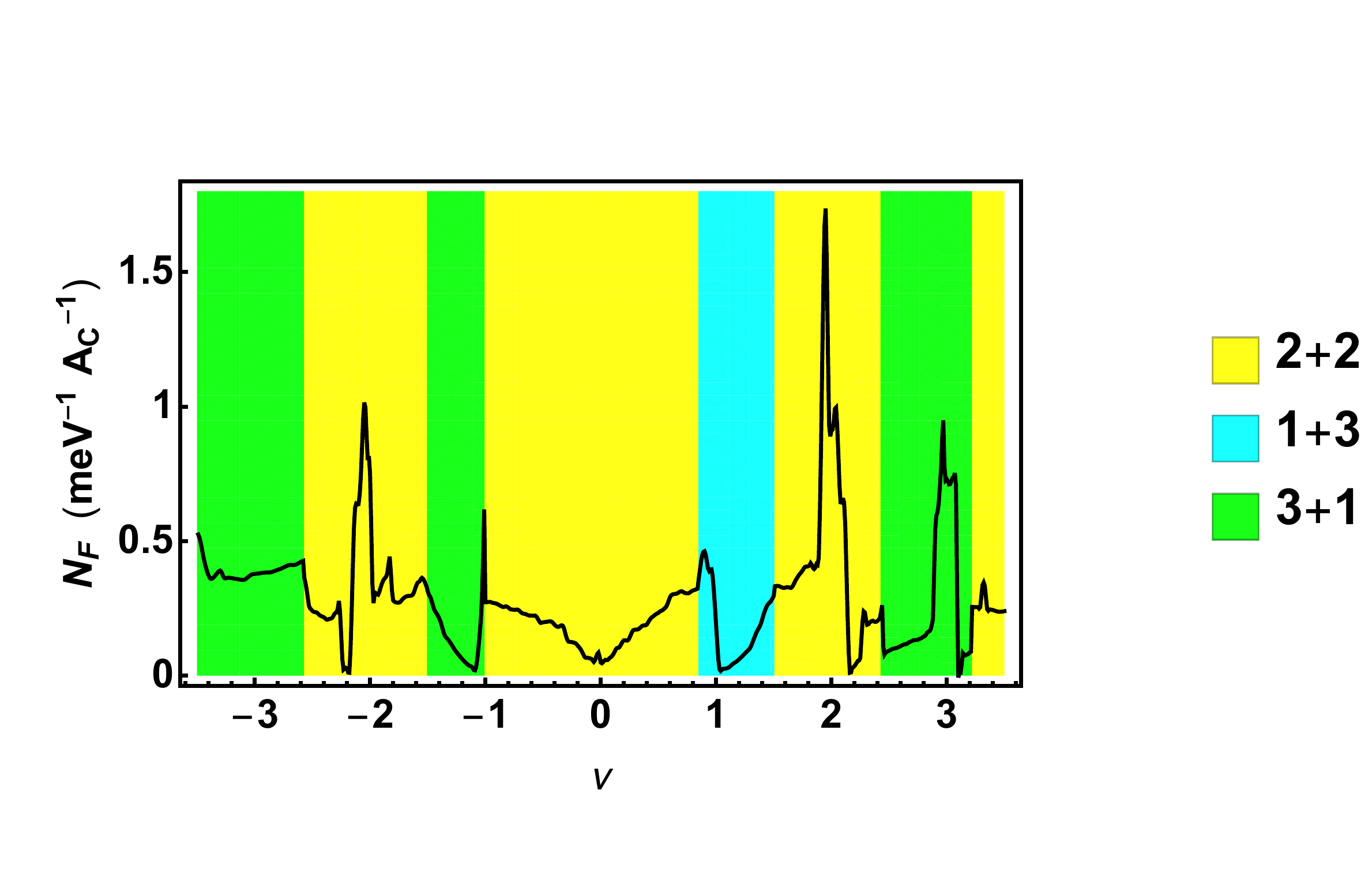}
\caption{DOS calculated at the Fermi level, $N_F$, as a function of $\nu$.
The different colors of the background specify the type of polarization stabilizing the GS at any filling.}
\label{NF}
\end{figure*}

Recent magnetotransport measurements \cite{cao_nat18,cao_fatemi_nat18,Yankowitz_science19} on TBG reported on the reduced degeneracy of the Landau levels from $8$ to $4$, which suggests that the four-fold spin/valley degeneracy of the bands of monolayer graphene is removed in TBG, in favor of polarization states\cite{ferro}.
We assume that charge is transferred from low occupancy bands to high occupancy bands. We consider three different configurations: i) charge is transferred from two sets (valence and conduction) of degenerate low occupancy bands to two sets of degenerate high occupancy bands,  labelled as 2+2 ii) one set of high occupancy bands, and three sets of degenerate low occupancy bands, 1+3 iii) three sets of degenerate high occupancy bands, and one set of low occupancy bands, 3+1. The population of the bands in each case, as function of the total charge, $\nu$, with $-4 \le \nu \le 4$, is given in the Table[\ref{polarization_filling_table}]. Results for self consistent solutions in these three cases are shown in the Fig.[\ref{polarized_bands_evolution}].

Comparison between the energies of the phases with broken symmetry discussed above is shown in Fig.[\ref{EGS}]. The differences in energies between different phases are of a few meV per unit cell, $A_C=\sqrt{3}L^2/2$. These small differences imply that interactions at the atomic scale can modify their relative stability. A change in the dielectric function of the environment can also change the lowest energy phase, as shown in Fig.[\ref{EGS_epsilon}]. Note that the Hartree-Fock approximation used here takes into account the increased susceptibility of the system when the Fermi energy is near a van Hove singularity\cite{SB18,KLK18,GS19,Letal18,LQYZLH19,cea_prb19,NIL19,IF19,FV19,CCC19,HWS20}.

A summary of the results is shown in Fig.[\ref{NF}]. We find that polarized phases have the lowest energy for the whole filling range.  The deformation in the band shape induced by the self consistent Hartree and Fock potentials leads to crossings between the polarized bands, so that the Fermi energy intersects some of the bands at all fillings. The resulting density of states (DOS) at the Fermi level, $N_F$, is also plotted in Fig.[\ref{NF}]. $N_F$ can be very low near integer fillings, suggesting a behavior similar to the pseudogap regime in the cuprate oxide superconductors.

Note that these results depend on the strength of the screening of the electrostatic potential. Near half filling, and for a high dielectric constant, $\epsilon\gtrsim 30$, the lattice polarized phase mentioned earlier has the lowest energy, as shown in the inset of Fig.[\ref{EGS_epsilon}]. This phase shows a gap at half filling, see Fig.[\ref{np_bands_evolution}].

The nature of the broken symmetry phases studied here, and their dependence on filling highlights a number of similarities and differences with other strongly correlated materials: 

i) Polarized, in valley or spin, phases are stable over a wide energy range, not only at integer fillings. Different types of polarization are possible. This scenario has been contemplated in\cite{zondiner_cm19}, and, most likely, it implies first order phase transitions, hysteresis, and electronic phase separation at mesoscopic scales. The stability of the 3+1 phase seems consistent with the observation of a magnetic phase in\cite{Setal19}, see Fig.[\ref{polarized_bands_evolution}, (c)].

ii) The polarized phases are typically gapless, as the electrostatic (Hartree) potential distorts the bands by a larger amount than the band splitting induced by the exchange term. The DOS at the Fermi level shows minima near integer fillings.

iii) Near half filling a sublattice polarized phase is favored for sufficiently small values of the interaction. This phase has been discussed in\cite{XM20,bultinck_cm19,liu_cm19,gonzalez_cm20}. This phase is stable within a range of densities around half filling. It shows a gap at half filling, and it is gapless at non integer fillings, see Fig.[\ref{np_bands_evolution}]. For higher interactions, we obtain a spin polarized metallic phase, see Fig.[\ref{polarized_bands_evolution}, (a)]. The existence of these competing phases may explain discrepancies in experimental observations\cite{cao_nat18,cao_fatemi_nat18,tomarken_prl19,xiaobo_nat19}.

iv) The energy differences between broken symmetry phases is of order of a few meV per unit cell. The balance between phases may be altered by interactions at the atomic scale\cite{atomic}.

v) In all cases considered here, and at all fillings, the combination of the electrostatic and exchange potentials leads to bandwidths of order $\sim e^2 / ( \epsilon L )$.

vi)  Phases where a continuous symmetry is broken lead to Goldstone modes. As in the case of the phase stiffness in a supercoducting phase\cite{HHPR19,XSLB19,JPLHT20}, the dispersion of these modes needs not be limited by the electronic bandwidth. These low energy modes may play a role in the temperature dependence of the conductivity\cite{Petal19,Cetal20}.

{\it Acknowledgements.}
This work was supported by funding from the European Commision, under the Graphene Flagship, Core 3, grant number 881603, and by the grants NMAT2D (Comunidad de Madrid, Spain),  SprQuMat and SEV-2016-0686, (Ministerio de Ciencia e Innovación, Spain).

\bibliography{Literature}

\begin{thebibliography}{56}
\bibitem [{c2()}]{c2}%
  \BibitemOpen
  \href@noop {} {}\bibinfo {note} {The spatial broken symmetry is a
  $\mathcal{C}_2$ rotation which exchanges the two sublattices within each
  layer.}\BibitemShut {Stop}%
\bibitem [{ato()}]{atomic}%
  \BibitemOpen
  \href@noop {} {}\bibinfo {note} {An atomic Hubbard term, $U$, projected onto
  the bands near half filling gives an energy scale $U_{eff} \sim U/N^2$, where
  $N \approx 10^4$ is the numbe of atoms in the unit cell. For $U \approx 2 -8
  eV$, one obtains a scale $\lesssim 1$ meV.}\BibitemShut {Stop}%
\end{thebibliography}


\clearpage
\onecolumngrid

\setcounter{equation}{0}
\setcounter{figure}{0}
\setcounter{table}{0}
\setcounter{page}{1}
\makeatletter
\renewcommand{\theequation}{S\arabic{equation}}
\renewcommand{\thefigure}{S\arabic{figure}}

\begin{center}
\Large Supplementary information for \\ Band structure and insulating states driven by the Coulomb interaction in twisted bilayer graphene
\end{center}

\section{Methods: the Hartree-Fock approximation within the continuum model of TBG}
\begin{figure}[h!]
\includegraphics[width=2.5in]{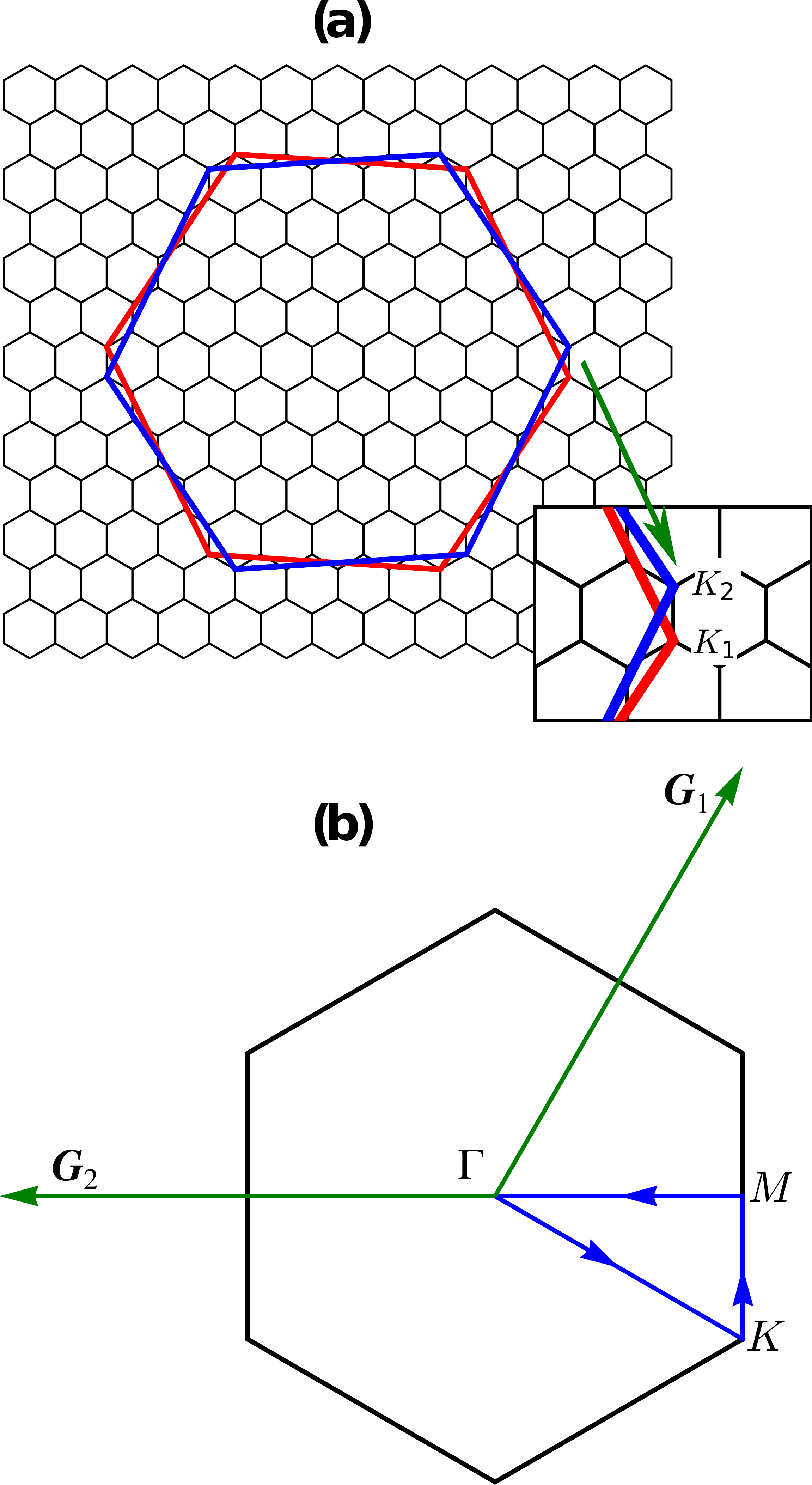}
\caption{
(a) Folding of the BZs of the twisted monolayers graphene.
The BZ of layer 1 (red hexagon) is rotated by $-\theta/2$, while that of the layer 2 (blue hexagon) by $\theta/2$.
The small black hexagons represent the mini-BZs forming the reciprocal moir\'e lattice.
In the inset: $K_{1,2}$ are the Dirac points of the twisted monolayers, which identify the corners of the mini-BZ.
(b) mini-BZ. $\vec{G}_{1,2}$ are the two reciprocal lattice vectors. The blue line shows the high symmetry circuit in the mini-BZ
used to compute the bands shown in the following.}
\label{BZ}
\end{figure}

Rotating two layers of graphene by a relative small angle, $\theta$,
gives rise to a moir\'e pattern.
The period of the moir\'e, $L=\frac{a}{2\sin(\theta/2)}$,
dramatically increases by reducing $\theta$,
where $a=2.46${\AA} is the lattice constant of graphene.
We describe the TBG within the low energy continuum model
considered in  Refs.\cite{santos_prl07,Bistritzer_pnas11,santos_prb12,koshino_prx18},
which is meaningful for sufficiently small angles,
so that an approximatively commensurate structure can be defined for any twist.
The moir\'e mini-BZ, resulting from the folding of the two BZs of each monolayer
(see Fig.\ref{BZ}(a)),
is generated by the two reciprocal lattice vectors:
\begin{equation}
\vec{G}_1=2\pi(1/\sqrt{3},1)/L\text{ and } \vec{G}_2=4\pi(-1/\sqrt{3},0)/L,
\end{equation}
 shown in green in Fig.~\ref{BZ}(b).

Let $K_\xi=\xi4\pi(1,0)/3a$ be the two Dirac points of the unrotated monolayer graphene, with $\xi=\pm1$.
For small twists, the coupling between the $K_+$ and $K_-$ valleys of the two monolayers
can be safely neglected, as the interlayer hopping has a long wavelength modulation.

The fermionic field operators of the TBG are  4-component Nambu spinors:
\bea
\Psi_{\xi\sigma}=\left(\psi_{\xi\sigma}^{A_1},\psi_{\xi\sigma}^{B_1},\psi_{\xi\sigma}^{A_2},\psi_{\xi\sigma}^{B_2}\right)^T,
\eea 
where $A,B$ and $1,2$ denote the sub-lattice and layer indices, respectively, and $\sigma$ is the spin index.
We introduce a relative twist $\theta$ between the two monolayers 
by rotating the layer $1$ by $-\theta/2$ and the layer $2$ by $\theta/2$.
Without loss of generality, we assume that in the aligned configuration, at $\theta=0$,
the two layers are $AA$-stacked.
In the continuum limit, the effective Hamiltonian of the TBG in a volume $\Omega$ can be generally written as
\cite{santos_prl07,Bistritzer_pnas11,santos_prb12,koshino_prx18}:
\bea\lb{HTBG}
\hat{H}_{TBG}=\sum_{\xi\sigma}\int_\Omega\,d^2\vec{r}\Psi_{\xi\sigma}^\dagger(\vec{r})
\begin{pmatrix}
H_{\xi1}&U_\xi(\vec{r})\\U_\xi^\dagger(\vec{r})&H_{\xi2}
\end{pmatrix}
\Psi_{\xi\sigma}(\vec{r}),
\eea
where
\bea
H_{\xi l}=\xi \hbar v_F \left(-i\vec{\nabla}-\xi K_l\right)\cdot
\vec{\tau}^\xi_{\theta_l}
\eea
is the Dirac Hamiltonian for the $\xi$ valley of layer $l$, $v_F=\sqrt{3}ta/(2\hbar)$ is the Fermi velocity,
$t$ is the hopping amplitude between localized $p_z$ orbitals at nearest neighbors carbon atoms,
$\theta_{1,2}=\mp\theta/2$,
$\vec{\tau}^\xi_{\theta_l}=e^{i\tau_z\theta_l/2}\left(\tau_x,\xi\tau_y\right)e^{-i\tau_z\theta_l/2}$,
$\tau_i$ are the Pauli matrices, and $K_l=4\pi\left(\cos\theta_l,\sin\theta_l\right)/(3a)$
are the Dirac points of the two twisted monolayers corresponding to the $\xi=+$ valley, which identify the corners of the mini-BZ shown in
Fig.~\ref{BZ}(a).
$U_\xi(\vec{r})$ is the inter layer potential, which is a periodic function in the moir\'e unit cell.
In the limit of small angles, its leading harmonic expansion is determined by only three reciprocal lattice vectors \cite{santos_prl07}:
$U_\xi(\vec{r})=U_\xi(0)+U_\xi\left(-\vec{G}_1\right)e^{-i\xi\vec{G}_1\cdot\vec{r}}+
U_\xi\left(-\vec{G}_1-\vec{G}_2\right)e^{-i\xi\left(\vec{G}_1+\vec{G}_2\right)\cdot\vec{r}}$,
where the amplitudes $U_\xi\left(\vec{G}\right)$ are given by:
\bea
U_\xi(0)&=&\begin{pmatrix}g_1&g_2\\g_2&g_1\end{pmatrix},\nn\\
U_\xi\left(-\vec{G}_1\right)&=&\begin{pmatrix}g_1&g_2e^{-2i\xi\pi/3}\\g_2e^{2i\xi\pi/3}&g_1\end{pmatrix},\\
U_\xi\left(-\vec{G}_1-\vec{G}_2\right)&=&\begin{pmatrix}g_1&g_2e^{2i\xi\pi/3}\\g_2e^{-2i\xi\pi/3}&g_1\end{pmatrix}.\nn
\eea
In the following we adopt the parametrization of the TBG
given in the Ref.\cite{koshino_prx18}: $\hbar v_F/a=2.1354$eV,
$g_1=0.0797$eV and $g_2=0.0975$eV.
The difference between $g_1$ and $g_2$, as described in\cite{koshino_prx18},
accounts for the inhomogeneous interlayer distance, which is minimum in the $AB/BA$ regions
and maximum in the $AA$ ones, or it can be seen as a model of a more complete treatment of lattice relaxation\cite{guinea_prb19}.
If we focus eg on the $\xi=+$ valley, then the Hamiltonian of Eq.~\pref{HTBG} hybridizes states of layer $1$ with momentum
$\vec{k}$ close to the Dirac point with the states of layer $2$ with momenta
$\vec{k},\vec{k}+\vec{G}_1,\vec{k}+\vec{G}_1+\vec{G}_2$.

\begin{figure}
\includegraphics[width=3.5in]{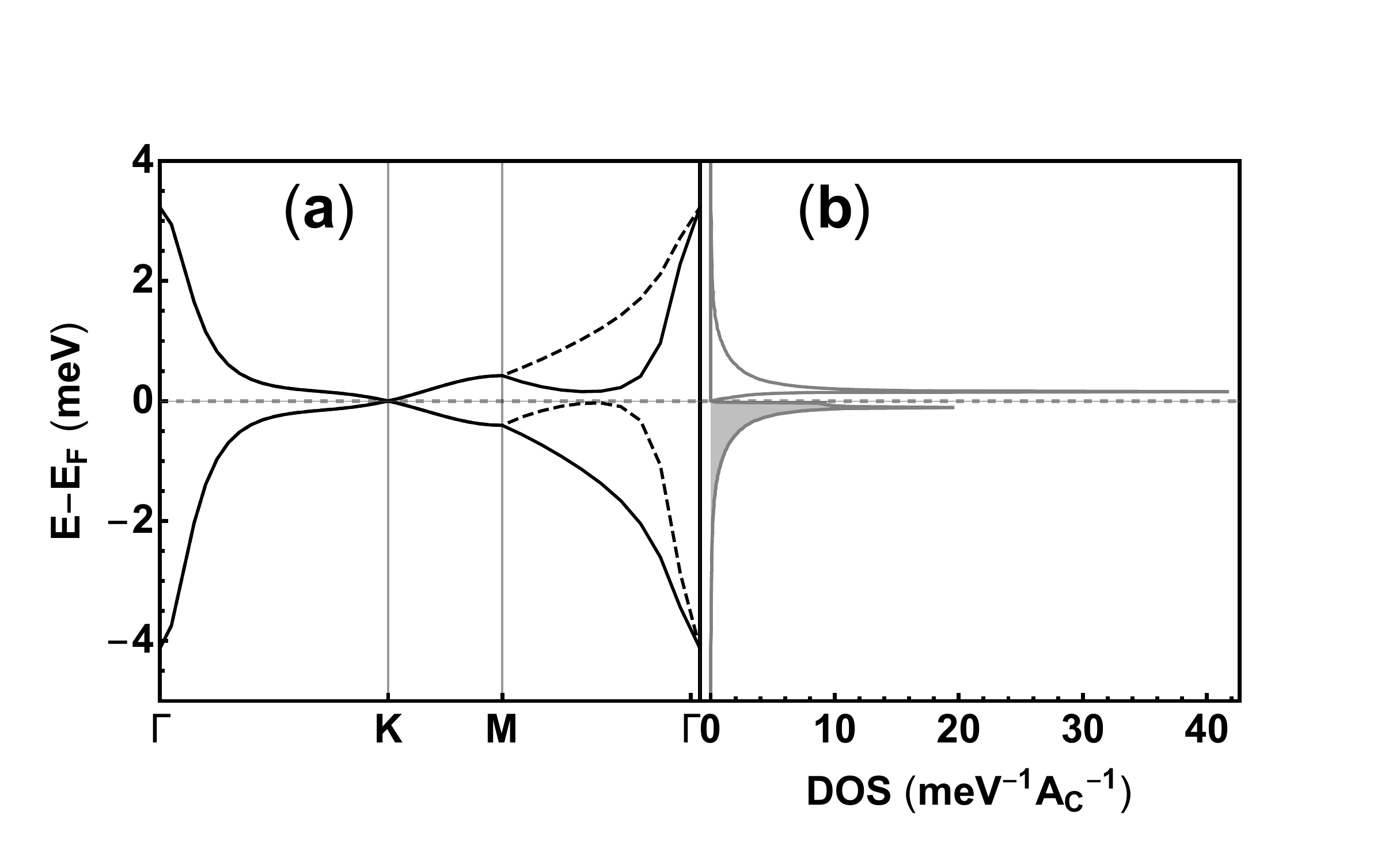}
\caption{
(a): mini-bands of the non-interacting TBG at CN, obtained for the twist angle $\theta=1.05^\circ$
and computed along the high symmetry circuit of the BZ denoted by the blue arrows in Fig.~\ref{BZ}(b).
The continuum and dashed black lines correspond to the $\xi=\pm$ valleys, respectively. They are related each other by the time-reversal symmetry, upon inverting $\vec{k}$ to $-\vec{k}$.
(b): DOS per moir\'e unit cell, normalized to 8.
}
\label{bands_koshino}
\end{figure}

In the absence of interactions, the band structure and the DOS per moir\'e unit cell of the mini-bands at CN are shown in Fig.~\ref{bands_koshino}, for $\theta=1.05^\circ$.
Here the origin of the energy axes has been set at CN, $E_F$ is the Fermi energy and
the bands are computed along the high symmetry circuit of the BZ denoted by the blue arrows in Fig.~\ref{BZ}(b).
The continuum and dashed black lines correspond to the $\xi=\pm$ valleys, respectively. They are related each other by the time-reversal symmetry, upon inverting $\vec{k}$ to $-\vec{k}$.
$A_C=L^2\sqrt{3}/2$ is the area of the moir\'e unit cell and the DOS is normalized to 8, accounting for two bands and four spin/valley flavors.
As deeply studied in the past literature, \cite{Bistritzer_pnas11,koshino_prx18,tarnopolsky_prl19}, these bands are characterized by a very narrow bandwidth, $\sim$meV,
and by an almost vanishing Fermi velocity as compared to that of monolayer graphene,
thus pinning the DOS at the two van Hove singularities in Fig.~\ref{bands_koshino}(b).

Next we introduce the Coulomb interaction, as described by the Hamiltonian:
\bea\lb{H_C}
\hat{H}_C=\frac{1}{2}\int_\Omega\,d^2\vec{r}d^2\vec{r}'\delta\hat{\rho}(\vec{r})v_C(\vec{r}-\vec{r}')\delta\hat{\rho}(\vec{r}'),
\eea
where $\delta\hat{\rho}(\vec{r})\equiv \hat{\rho}(\vec{r})-\rho_{CN}(\vec{r})$ is the quantum operator associated to the density fluctuations,
$\hat{\rho}(\vec{r})=\sum_{ \mu}\Psi^\dagger_\mu(\vec{r})\Psi_\mu(\vec{r})$ is the density operator,
$\mu=(\xi,\sigma)$ being the generalized valley/spin index,
$\rho_{CN}(\vec{r})$ is the average density corresponding to the non-interacting TBG at CN,
and $v_C(\vec{r})$ is the Coulomb potential.
In the following, we assume that the Coulomb interaction is screened by
a double metallic gate,
as described by the Fourier envelope:
$v_C(\vec{q})\equiv\int\,d^2\vec{r}v_C(\vec{r})e^{-i\vec{q}\cdot\vec{r}}=\frac{2\pi e^2}{\epsilon |\vec{q}|}\tanh\left(|\vec{q}|d\right)$,
where $e$ is the electron charge,
$\epsilon$ the dielectric constant of the environment and $d$ the distance of the sample from the gate.
We set: $\epsilon=10$ and $d=40$nm, which are realistic values in the experiments.

At mean-field level, the Hamiltonian $\hat{H}_C$ is replaced by:
\bea\label{HCMF}
\hat{H}_C\rightarrow \hat{H}^{MF}_C=\hat{H}_H+\hat{H}_F+E_0,
\eea 
where:
\begin{subequations}\lb{HFterms}
\bea\lb{Hartree}
\hat{H}_H=\sum_{i\mu}\int_\Omega\,d^2\vec{r}\psi^{i,\dagger}_\mu(\vec{r})\psi^{i}_\mu(\vec{r}) V_H(\vec{r})
\eea
is the Hartree Hamiltonian, $V_H(\vec{r})=\int_\Omega\,d^2\vec{r}'v_C(\vec{r}-\vec{r}')\left\langle\delta\hat{r}(\vec{r}')\right\rangle$ being the local Hartree potential,
\bea\lb{Fock}
\hat{H}_F=\sum_{ij\mu}\int_\Omega\,d^2\vec{r}\,d^2\vec{r}'\psi^{i,\dagger}_\mu(\vec{r})V^{ij}_{F,\mu}(\vec{r},\vec{r}')\psi^{j}_\mu(\vec{r}')
\eea
is the Fock Hamiltonian, $V^{ij}_{F,\mu}(\vec{r},\vec{r}')=-\left\langle \psi^{j,\dagger}_\mu(\vec{r}')   \psi^{i}_\mu(\vec{r})\right\rangle v_C(\vec{r}-\vec{r}') $ being the non-local Fock potential,
and
\bea
E_0&=&-\frac{1}{2}\left[\left\langle   \hat{H}_H+ \hat{H}_F \right\rangle\
+\int_\Omega\,d^2\vec{r}\rho_{CN}(\vec{r})V_H(\vec{r})\right]
\eea
is the zero point energy, which is required to avoid double counting of the total energy at mean-field level.
\end{subequations}
The mean-field Hamiltonian for the interacting TBG is then:
\bea\label{HMF}
\hat{H}^{MF}=\hat{H}_{TBG}+\hat{H}_C^{MF},
\eea
which we diagonalize self-consistently, by computing the quantum averages of the Eqs. \pref{HFterms} by means of $\hat{H}^{MF}$
and iterating until convergence.
It's worth noting that this procedure is equivalent to minimize the GS energy of $\hat{H}^{MF}$.
In order to diagonalize $\hat{H}^{MF}$, we exploit the Bloch's theorem, by expressing the eigenfunctions in the basis of Bloch's plane waves defined on the moir\'e:
\bea\lb{eigenstates}
\ket{\vec{k},\alpha,\mu}=\sum_{\vec{G}i}\phi_{\vec{k}+\vec{G},\alpha,\mu,i}\ket{\vec{k}+\vec{G},\mu,i},
\eea 
where $\vec{k}\in$mBZ, the $\vec{G}$'s are reciprocal lattice vectors, $\alpha$ is the band index and $\phi_{\vec{k}+\vec{G},\alpha,\mu,i}$
are numerical eigenvectors normalized according to:
$\sum_{i\vec{G}}\phi^*_{\vec{k}+\vec{G},\alpha,\mu,i}\phi_{\vec{k}+\vec{G},\alpha',\mu,i}=\delta_{\alpha \alpha'}$.
Upon using the Eq. \pref{eigenstates} to evaluate the Hartree and Fock potentials,
the matrix elements of the Eqs. \pref{HFterms} can be written in the Bloch's basis as: 
\begin{subequations}\lb{matrixelements}
\bea\lb{Hartree_mat_el}
\bra{\vec{k}+\vec{G},\mu,i}\hat{H}_H\ket{\vec{k}+\vec{G}',\mu',i'}=
\delta_{ii'}\delta{\mu\mu'}\frac{v_C(\vec{G}-\vec{G}')}{\Omega}
\times\nn\\
\times\sum_{\vec{k}'\vec{G}''}\sum_{\alpha \mu'' i''}
\phi_{\vec{k}'+\vec{G}''+\vec{G},\alpha,\mu'',i''}
\phi^*_{\vec{k}'+\vec{G}''+\vec{G}',\alpha,\mu'',i''}\equiv\delta_{ii'}\delta{\mu\mu'} V_H\left(\vec{G}-\vec{G}'\right),
\eea
where the sum over the band index, $\alpha$, runs over all the occupied states counted from CN, and $V_H\left(\vec{G}\right)$ is noting but the Fourier transform of the Hartree potential, $V_H\left(\vec{r}\right)$, evaluated in $\vec{G}$.
\bea\lb{Fock_mat_el}
\bra{\vec{k}+\vec{G},\mu,i}\hat{H}_F\ket{\vec{k}+\vec{G}',\mu',i'}&=&\nn\\
-\delta_{\mu\mu'}\sum_{\vec{k}'\vec{G}''\alpha}\frac{v_C(\vec{k}-\vec{k}'-\vec{G}'')}{\Omega}
\phi_{\vec{k}'+\vec{G}''+\vec{G},\alpha,\mu,i}\phi^*_{\vec{k}'+\vec{G}''+\vec{G}',\alpha,\mu,i'}&\equiv&
\delta_{\mu\mu'}
 \Sigma_F^\mu ( \vec{k} + \vec{G} , i ; \vec{k} + \vec{G}' , i' ),
\eea
\end{subequations}
where $\alpha$ runs over all the occupied states above a given threshold.
In the present context, we set this threshold to the lowest energy of the mini-bands in the middle of the spectrum,
meaning that we are neglecting the contribution of the bulk bands. However, including other bands might affect quantitatively the results.
Because the Eqs. \pref{matrixelements} express the matrix elements in terms of the energy levels and of the corresponding eigenfunctions, $\phi$,
they completely define the self-consistent problem.

The main contributions of the long-range interaction are expected to come from small momenta.
Therefore, we only consider the matrix elements of the Hartree potential, Eq. \pref{Hartree_mat_el},
with $\vec{G}-\vec{G}'$ belonging to the first star of reciprocal lattice vectors: $\pm\vec{G}_1,\pm\vec{G}_2,\pm\left(\vec{G}_1+\vec{G}_2\right)$.
Concerning the matrix elements of the Fock potential, Eq. \pref{Fock_mat_el},
for each external momentum $\vec{k}$ we truncate the sum over $\vec{k}'$ and $\vec{G}''$ so that: $\vec{k}-\vec{k}'-\vec{G}''$ belongs to the BZ.
We checked that including larger momenta affects negligibly the results.

Finally, the energy of the GS, as following from the Eq.s \pref{HCMF}-\pref{HMF}, is given by:
\bea
E_{GS}=\sum_{\vec{k}\alpha\mu}\varepsilon_{\vec{k}\alpha\mu}+E_0,
\eea
where $\varepsilon_{\vec{k}\alpha\mu}$ are the single-particle energies
and the sum over $\alpha$ runs over all the occupied states.

\section{Evolution of the band structure as function of filling}

\begin{figure}
\includegraphics[width=3.in]{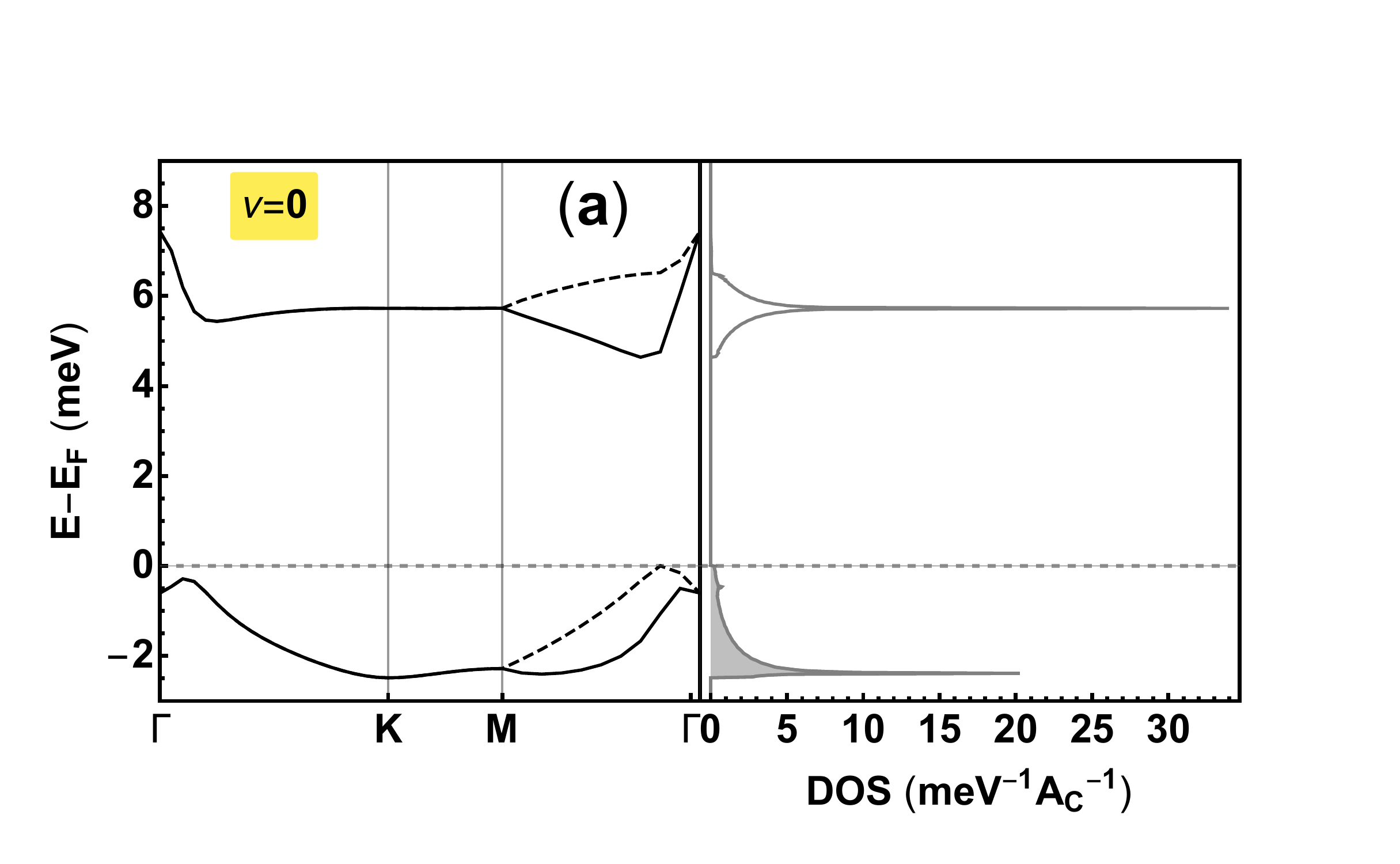}
\includegraphics[width=3.in]{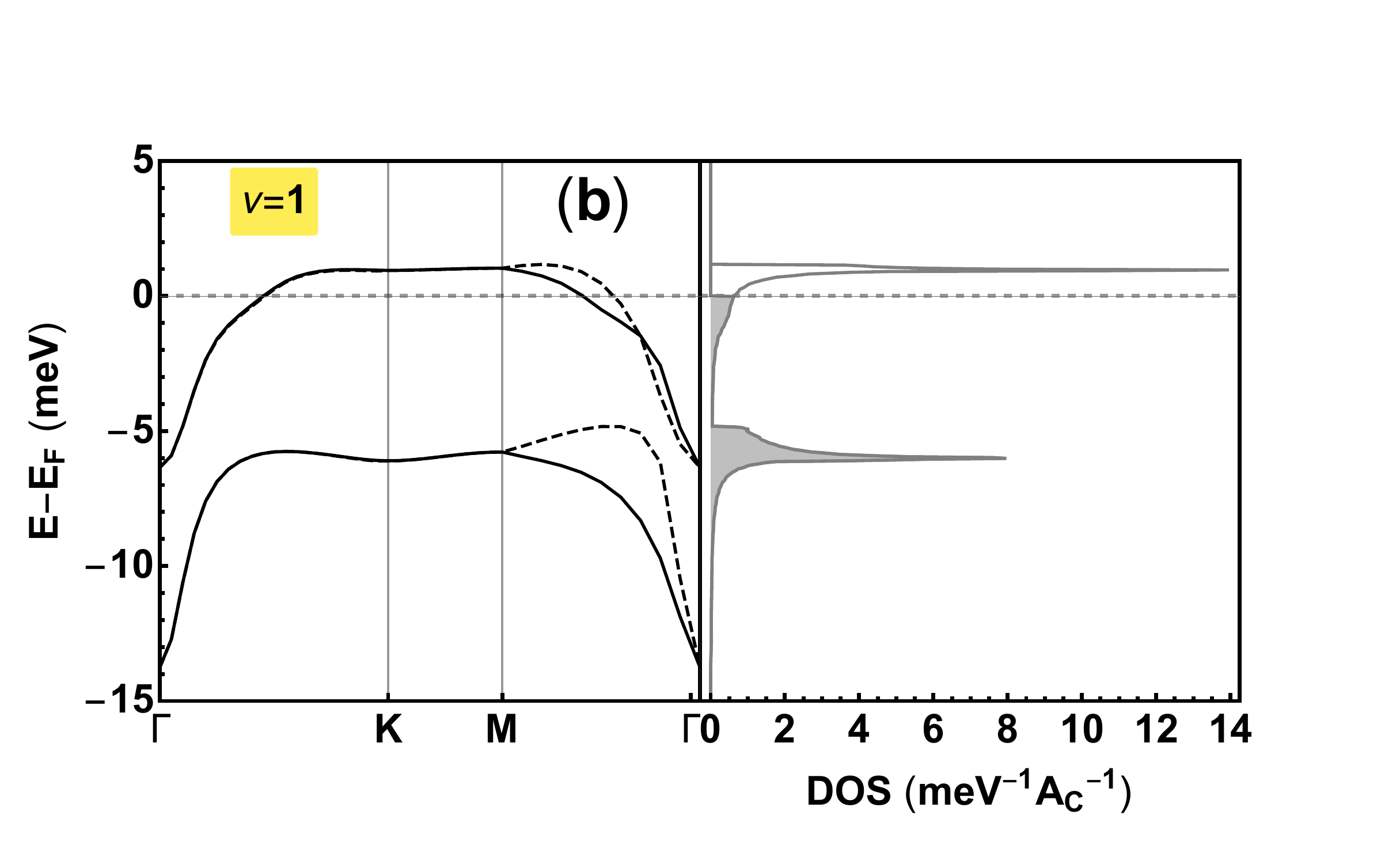}
\includegraphics[width=3.in]{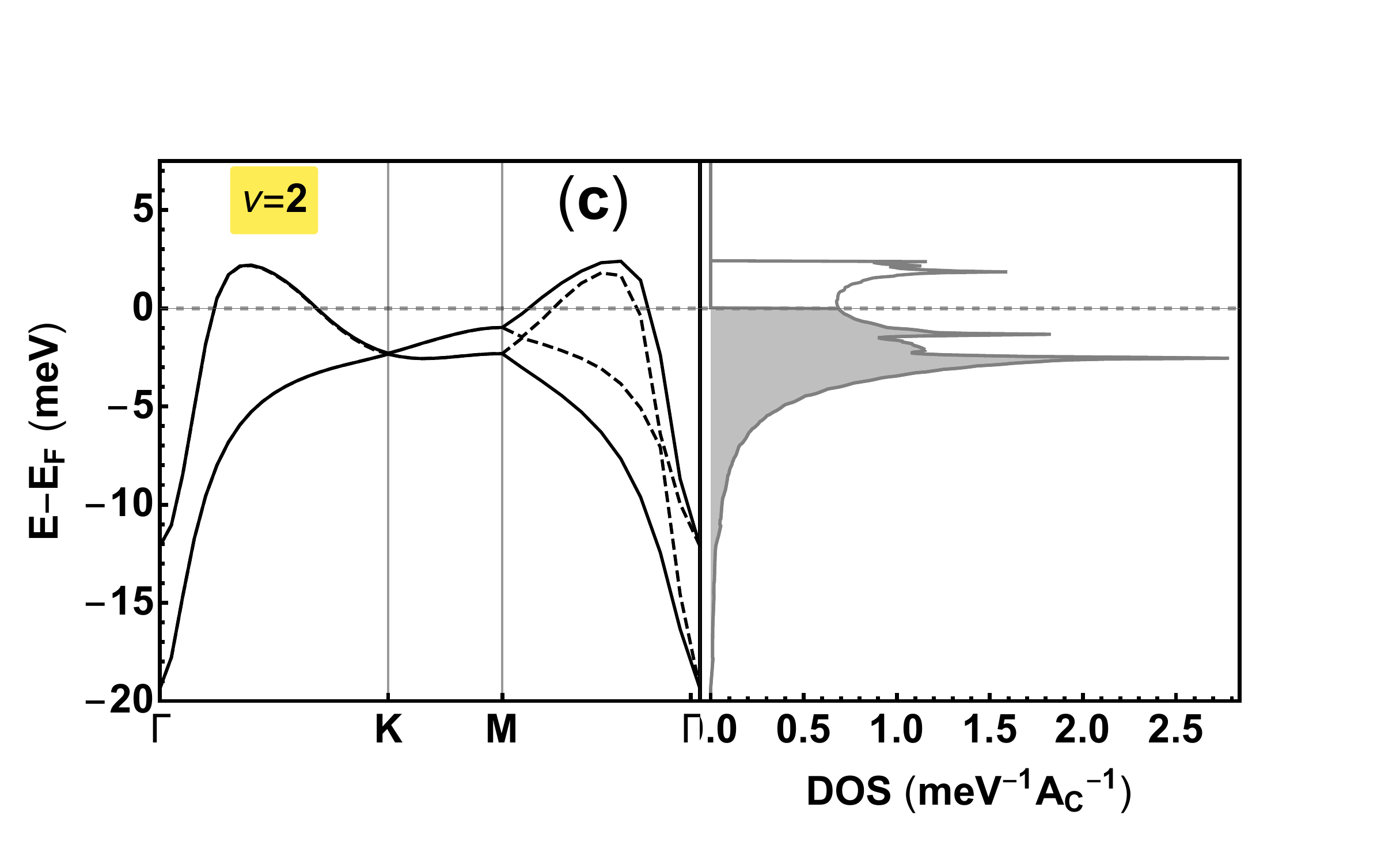}
\caption{Band structure and DOS corresponding to the non-polarized GS,
obtained within the Hartree-Fock approximation at the twist angle $\theta=1.05^\circ$
and filling: $\nu=0(a),1(b),2(c)$.}
\label{bands_np}
\end{figure}

Fig. \ref{bands_np} shows the band structure and DOS of the mini-bands of the interacting TBG at $\theta=1.05^\circ$,
for the non-polarized GS at positive filling of the conduction bands: $\nu=0,1,2$,
corresponding to CN, one and two electrons per moir\'e unit cell, respectively.
Each band is four-fold degenerate, so that the spin/valley flavors are equally occupied.
At $\nu=0$, only the lower band is filled, the $\mathcal{C}_2$ symmetry is broken and the Fermi surface (FS) is fully gapped.
For this choice of parameters, the width of the gap is $\sim 5$meV, comparable to the overall bandwidth.
At $\nu=1$, one quarter of the upper band is filled, the $\mathcal{C}_2$ symmetry is still broken, but the FS exists.
At $\nu=2$, the $\mathcal{C}_2$ symmetry is completely restored. 

\begin{figure*}
\centering
\includegraphics[width=7.in]{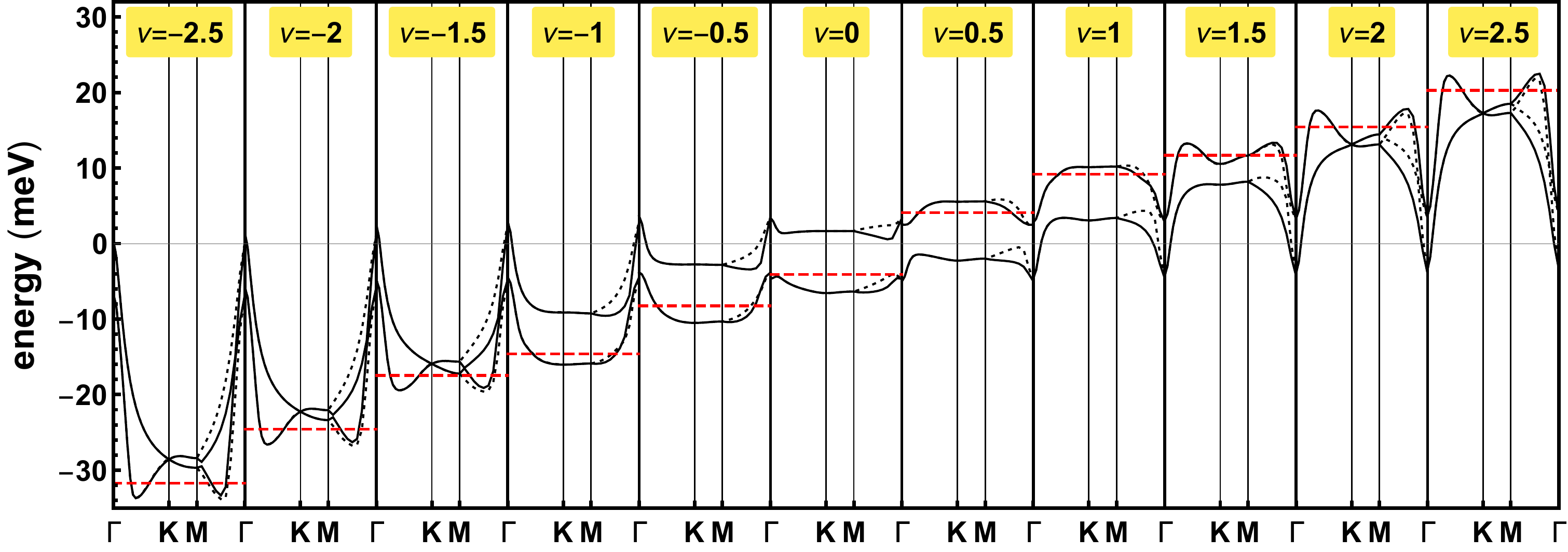}
\caption{Evolution of the band structure upon varying the filling, for $-2.5\leq\nu\leq2.5$.
The origin of the energy axis is the same for each panel.
The values of the Fermi energy are displayed by the horizontal red lines.}
\label{np_bands_evolutionSI}
\end{figure*}

The evolution of the band structure corresponding to the non-polarized solution is shown in Fig. \ref{np_bands_evolutionSI}, for $-2.5\leq\nu\leq2.5$.
The values of the Fermi energy are displayed by the horizontal red lines.
Here the origin of the energy axis is the same for each panel, so that curves corresponding to different values of $\nu$ can be compared each other.
As it can be seen, solutions breaking the $\mathcal{C}_2$ symmetry are not stable for $\nu>1.5$ and $\nu\le -1.5$.

Two features of the band structure are worth to be further noticed:
i) the lack of particle-hole symmetry, so that the bands are not symmetric upon inverting $\nu$ to $-\nu$;
ii) the bands are rigid at the $\Gamma$ point of the BZ. This is a consequence of the fact that the charge density of the TBG evaluated in $\Gamma$ is almost homogeneous as compared to the other high symmetry points of the BZ, as it has been already emphasized in the Refs.\cite{rademaker_prb18,Guinea_pnas18,cea_prb19}.

\begin{figure*}
\centering
\includegraphics[width=5.in]{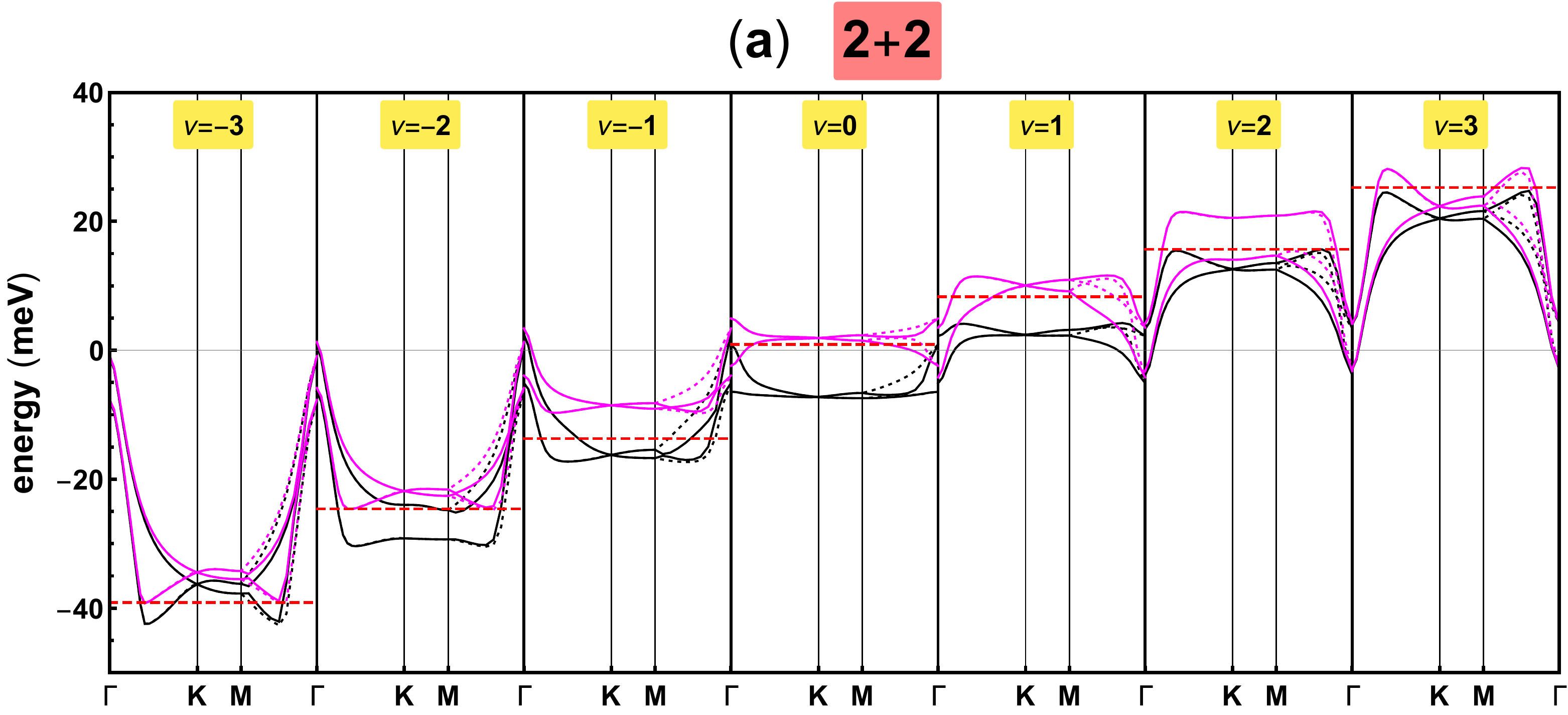}
\includegraphics[width=5.in]{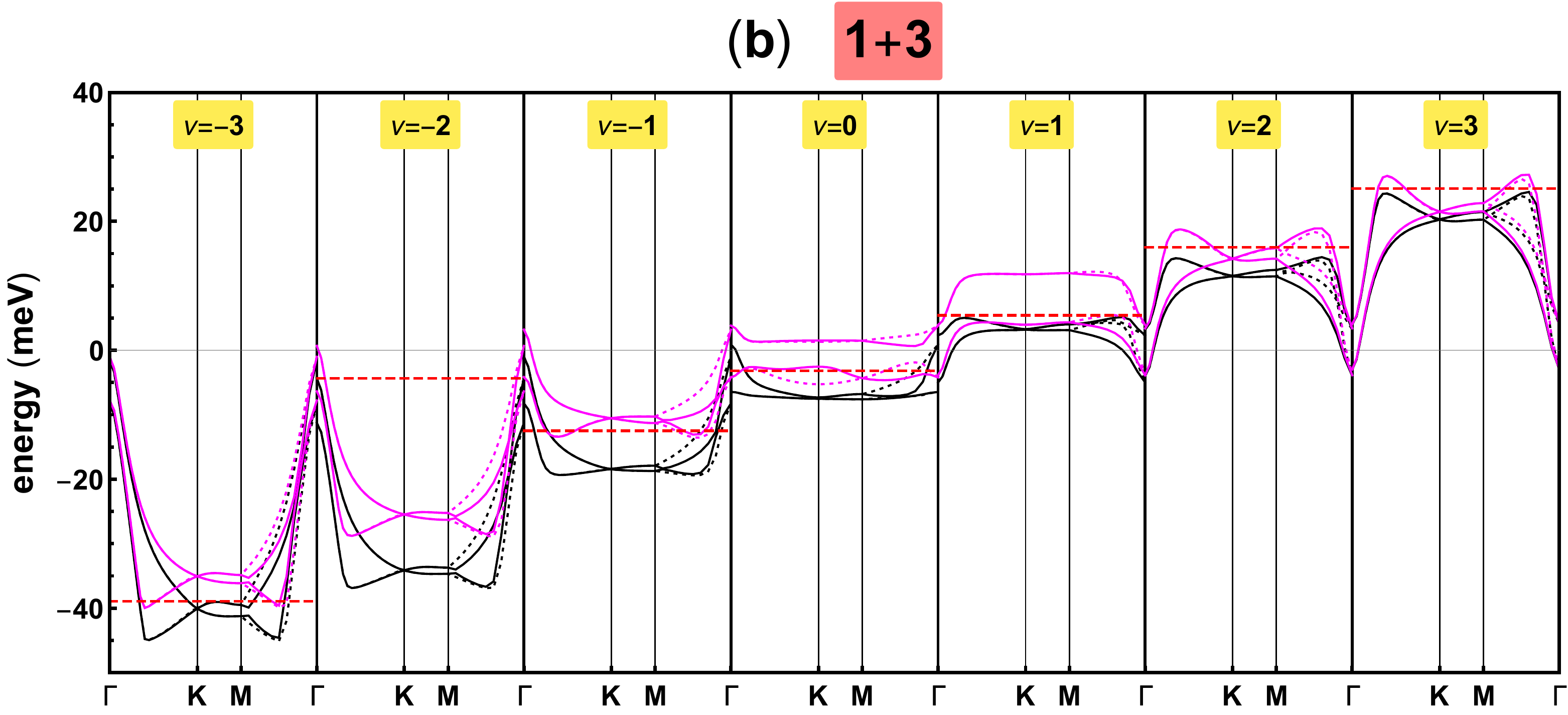}
\includegraphics[width=5.in]{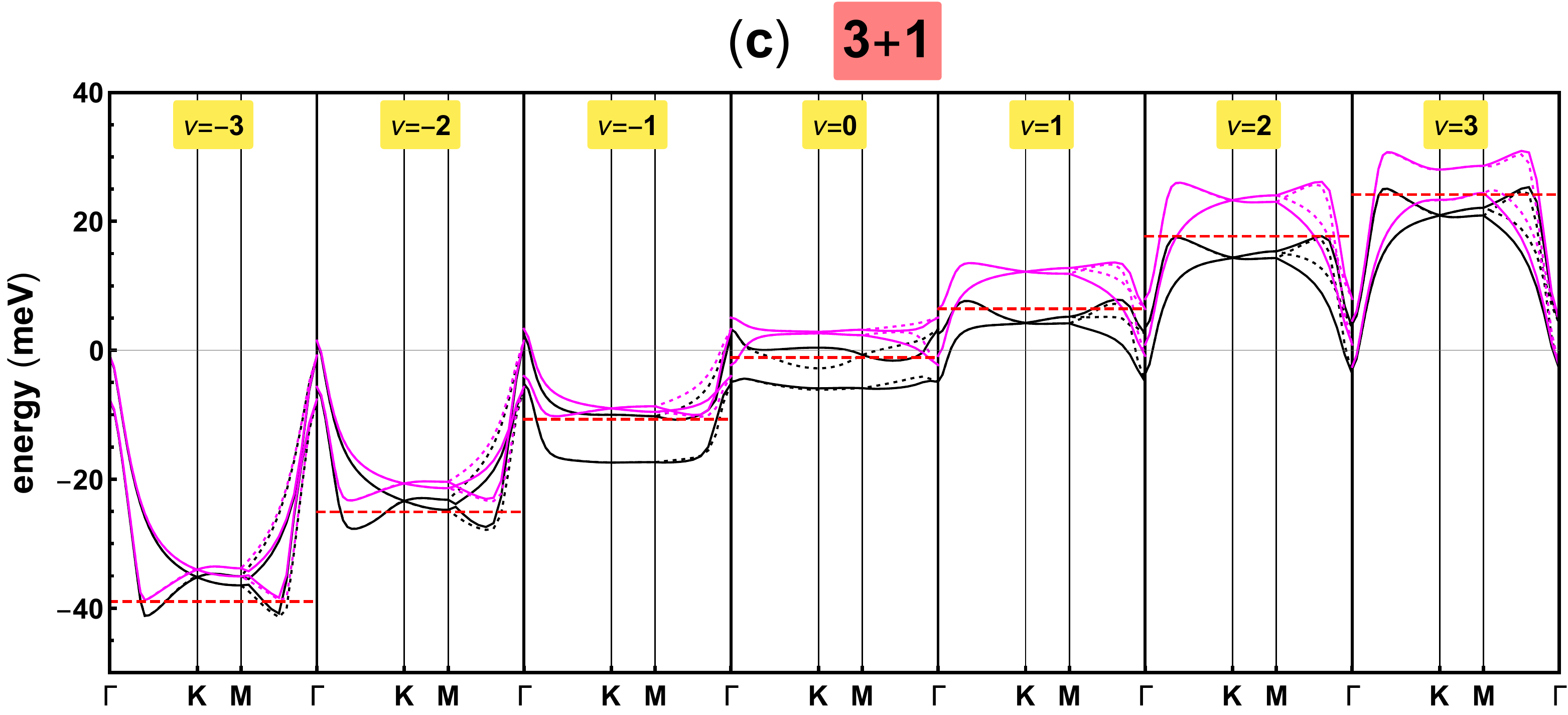}
\caption{
Band structure obtained at integer fillings, $-3\le\nu\le3$,
for the polarized solutions: $2+2$ (a), $1+3$ (b) and $3+1$ (c).
The high and low occupancy sets of bands are represented by the black and magenta lines, respectively.
}
\label{polarized_bands_evolutionSI}
\end{figure*}

\begin{figure}
\centering
\includegraphics[width=4.in]{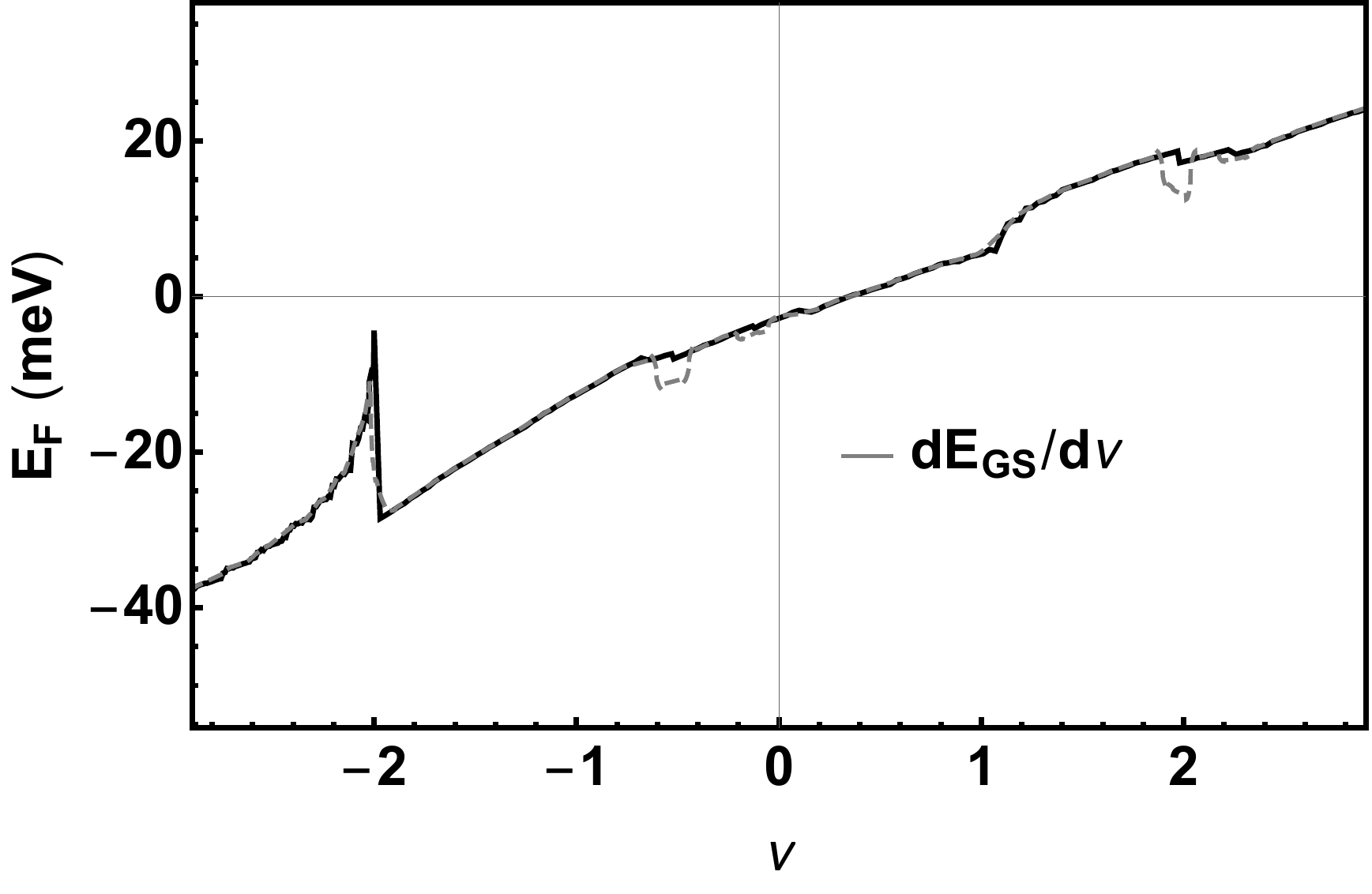}
\caption{
Fermi energy as a function of the filling, obtained in the $1+3$ solution.
The continuum black line shows the value of $E_F$ as obtained from the highest energy of the occupied states, while the dashed gray line represents $E_F$ computed as the derivative of the GS energy with respect to $\nu$.
}
\label{EF_vs_nu}
\end{figure}

Fig. \ref{polarized_bands_evolutionSI} shows the evolution of the band structure of the polarized solutions: $2+2$ (a), $1+3$ (b) and $3+1$ (c),
for integer fillings: $-3\le\nu\le3$. The high and low occupancy sets of bands are represented by the black and magenta lines, respectively. Solutions breaking the $\mathcal{C}_2$ symmetry occur at $\nu=\pm2$ for the $2+2$ solution, $\nu=1$ for the $1+3$ solution and $\nu=-1,3$ for the $3+1$ solution. Interestingly, the solution $1+3$ does not break $\mathcal{C}_2$ at $\nu=-3$, in contrast to what expected. We argue that here the $\mathcal{C}_2$ symmetry breaking is prevented by the small value of the interaction. Furthermore, it's worth noting that, in the $1+3$ solution at $\nu=-2$, the empty low occupancy bands mostly stay below the Fermi level. At this filling the Fermi energy, which we define as the highest energy of the occupied states, decreases upon increasing the filling, implying a negative compressibility. This behavior can be better seen in the Fig. \ref{EF_vs_nu}, showing $E_F$ as a function of $\nu$ and displaying a jump-like discontinuity at $\nu=-2$. The dashed line in the Fig. \ref{EF_vs_nu} represents the value of $E_F$ as obtained from the derivative of the GS energy with respect to $\nu$, which indeed matches quite well the curve of $E_F$ computed as described above. However, this anomalous behavior is actually not very meaningful in the present context, as the solution $1+3$ turns out to not be stable close to $\nu=-2$, as emphasized by the Figs. [4] and [6] of the main text.  

\section{inter-valley coherent phase}

A possible phase with broken symmetry in a twisted graphene bilayer at half filling is a phase which shows a sublattice polarization, breaks ${\mathcal C}_2$ symmetry, and opens a gap at the Dirac points. Hartree Fock calculations which include the long range electrostatic interaction  suggest this possibility\cite{XM20,bultinck_cm19}.

At half filling, the Hartree potential vanishes. The Fock term induces an interaction between electrons with the same flavor, spin and valley. Hence, a Hartree Fock approximation leads to four independent and equivalent hamiltonians. The sublattice polarization can have two values of opposite signs. A possible global ground state combines solutions for each electron flavor. Finally, rotations in spin and flavor space allows for an infinite number of solutions.

It has been argued that this degeneracy is broken by terms which are odd in the sublattice index, even if only long range interactions are considered\cite{bultinck_cm19}. The optimal ground state is given by the hybridization of solutions in different values with different sublattice polarizabilities.

In the following, we investigate further this issue. As in\cite{bultinck_cm19}, we restrict the electronic states to the two narrow bands near half filling obtained in a calculation without the interactions. We study phases obtained by hybridizing solutions for opposite valleys with opposite sublattice polarization. Other combinations can be studied in the same way. Neglecting the spin index, the wavefunctions that we consider can be written as $| \vec{k} , V , b \rangle$, where $\vec{k}$ labels the momentum, $V = K , K'$ labels the valley, and $b = v , c$ labels the band. In the absence of interactions, the valence band, $v$, is occupied, and the conduction band, $c$ is empty.

We assume that the exchange term is larger than the non interacting bandwidth. The exchange favors a large overlap between occupied states. 
We define the sublattice operator $\sigma_z = \sigma_z^1 + \sigma_z^2$ where superindex $1 , 2$ refer to the layer.  We also define the matrix $\sigma_{i,j}^{\vec{k}} = \langle , \vec{k}, i  , V | \sigma_z | \vec{k} , j , V \rangle$, $i = v , c$ , $V = K , K'$. This matrix has two eigenvalues of opposite sign, which depend on $\vec{k}$. The eigenvectors define superpositions of states in the valence and conduction bands with maximum sublattice polarization. We label these states as  $| \vec{k} , S , V \rangle$, with $S = A , B$ being the preferred sublattice polarization. Note that, unless the interlayer hoppings $V_{AA} = V_{BB}$ vanish, the sublattice polarization is less than unity.

 We that the exchange potential favors a ground state wavefunction formed from the sublattice polarized states  $| \vec{k} , S , V \rangle$. We consider a state made from a superposition of valleys $K$ and $K'$ with opposite polarizations, $S = A , B$,
\begin{align}
| \vec{k} \rangle &= \cos ( \theta_{\vec{k}} ) | \vec{k} , K , A \rangle + \sin ( \theta_{\vec {k}} ) e^{i \phi_{\vec{k}}} | - \vec{k} , K' , B \rangle
\end{align}
where $\theta_{\vec{k}}$ and $\phi_{\vec{k}}$ minimize the exchange energy.

The value of the exchange energy is
\begin{align}
E_{exch} &= - \frac{1}{2} \sum_{\vec{k} \in \Omega_{BZ}} \langle \vec{k} | \Sigma_{ex} ( \vec{k} ) | \vec{k} \rangle = \nonumber \\ &=  - \frac{1}{2} \sum_{\vec{k} , \vec{k}' \in \Omega_{BZ}} v_C ( \vec{k} - \vec{k}' ) | \langle \vec{k} | \vec{k}' \rangle |^2 = \nonumber  \\
&=  - \frac{1}{2} \sum_{\vec{k} , \vec{k}' \in \Omega_{BZ}} v_C ( \vec{k} - \vec{k}' ) \times \nonumber \\
&\times \left| \cos ( \theta_{\vec{k}} ) \cos ( \theta_{\vec{k}'} ) \langle \vec{k} , K , A | \vec{k}' , K , A \rangle + \right. \nonumber \\
&+ \left.  \sin ( \theta_{\vec{k}} ) \sin ( \theta_{\vec{k}'} ) e^{i ( \phi_{\vec{k}} - \phi_{\vec{k}'} )} \langle - \vec{k} , K' , B | - \vec{k}' ,K' , B \rangle \right|^2
\end{align}
where the Coulomb potential is $v_C ( \vec{q} ) = ( 2 \pi e^2 ) / ( \epsilon | \vec{q} | ) \times \tanh ( | \vec{q} | d )$, and $\epsilon$ is the dielectric constant of the environment. We neglect Umklapp processes. 

The value of the exchange energy depends on the absolute values of the complex numbers $\langle \vec{k} , K , A | \vec{k}' , K , A \rangle$ and $\langle - \vec{k} , K' , B | - \vec{k}' ,K' , B \rangle$, and on their relative phase. The dependence of the relative phase can be canceled by the value of $\phi_{\vec{k}} - \phi_{\vec{k}'}$. Then, the exchange energy is 
\begin{align}
E_{exch}  &=  - \frac{1}{2} \sum_{\vec{k} , \vec{k}' \in \Omega_{BZ}} v_C ( \vec{k} - \vec{k}' ) | \langle \vec{k} | \vec{k}' \rangle |^2 = \nonumber  \\
&=  - \frac{1}{2} \sum_{\vec{k} , \vec{k}' \in \Omega_{BZ}} v_C ( \vec{k} - \vec{k}' ) \times \nonumber \\
&\times \left| \cos ( \theta_{\vec{k}} ) \cos ( \theta_{\vec{k}'} ) | \langle \vec{k} , K , A | \vec{k}' , K , A \rangle | + \right. \nonumber \\
&+ \left.  \sin ( \theta_{\vec{k}} ) \sin ( \theta_{\vec{k}'} ) | \langle - \vec{k} , K' , B | - \vec{k}' ,K' , B \rangle | \right|^2
\end{align}
The $\mathcal{C}_2$ symmetry of the non interacting hamiltonian implies the equivalence $ \{ A , K \} \leftrightarrow \{ B , K' \}$ and $\{ B , K \} \leftrightarrow \{ A , K' \}$. Hence,
\begin{align}
 | \langle \vec{k} , K , A | \vec{k}' , K , A \rangle | &= | \langle - \vec{k} , K' , B | - \vec{k}' ,K' , B \rangle | = O_{\vec{k} - \vec{k}'}
\end{align}
and
\begin{align}
E_{exch}  &=   - \frac{1}{2} \sum_{\vec{k} , \vec{k}' \in \Omega_{BZ}} v_C ( \vec{k} - \vec{k}' ) \times  \cos^2 ( \theta_{\vec{k}} - \theta_{\vec{k}'}  )  \times O_{\vec{k} - \vec{k}'}^2
\end{align}
The lowest energy Hartree-Fock solution takes place for $\theta_{\vec{k}} = \theta_{\vec{k}'} = \theta$. The value of $\theta$ is arbitrary, so that the long range electrostatic interactions, treated within the Hartree Fock approximation does not favor a specific correlation between the occupancies of the two sublattices. Interactions at the atomic scale, however, will favor a phase in which the two sublattices have equal occupation, over the phase where all the charge resides in the same sublattice\cite{AF06}.
\end{document}